\documentclass[prc,showpacs,showkeys,superscriptaddress,twocolumn]{revtex4}
\usepackage{amsfonts}
\usepackage{amsbsy}
\usepackage{mathrsfs}
\usepackage{graphicx}
\def\cal{\mathcal}
\def\lsim{\mathrel{\rlap{
\lower4pt\hbox{\hskip-3pt$\sim$}}
    \raise1pt\hbox{$<$}}}     %less than approx. symbol
\def\gsim{\mathrel{\rlap{
\lower4pt\hbox{\hskip-3pt$\sim$}}
    \raise1pt\hbox{$>$}}}     %greater than or approx. symbol
\def\vec#1{\mbox{\boldmath $#1$}}
\def\scr#1{\mbox{\scriptsize #1}}
\begin{document}
\title{Lattice QCD Constraints on Hybrid and Quark Stars}
\author{Yu.B.~Ivanov}\thanks{e-mail: Y.Ivanov@gsi.de}
\affiliation{Gesellschaft
 f\"ur Schwerionenforschung mbH, Planckstr. 1,
D-64291 Darmstadt, Germany}
\affiliation{Kurchatov Institute, Kurchatov
sq. 1, Moscow 123182, Russia}
\author{A.S.~Khvorostukhin}\thanks{e-mail: hvorost@thsun1.jinr.ru}
\affiliation{Joint Institute for Nuclear Research,
RU-141980 Dubna, Moscow Region, Russia}
\author{E.E.~Kolomeitsev}\thanks{e-mail: kolomeitsev@physics.umn.edu}
\affiliation{School of Physics and Astronomy, University of
Minnesota, %116 Church St. SE,
Minneapolis, MN 55455, USA}
\author{V.V.~Skokov}\thanks{e-mail: V.Skokov@gsi.de}
\affiliation{Gesellschaft
 f\"ur Schwerionenforschung mbH, Planckstr. 1,
D-64291 Darmstadt, Germany}
\affiliation{Joint Institute for Nuclear Research,
RU-141980 Dubna, Moscow Region, Russia}
\author{V.D.~Toneev}\thanks{e-mail: V.Toneev@gsi.de}
\affiliation{Gesellschaft
 f\"ur Schwerionenforschung mbH, Planckstr. 1,
D-64291 Darmstadt, Germany}
\affiliation{Joint Institute for Nuclear Research,
RU-141980 Dubna, Moscow Region, Russia}
\author{D.N.~Voskresensky}\thanks{e-mail: D.Voskresensky@gsi.de}
\affiliation{Gesellschaft
 f\"ur Schwerionenforschung mbH, Planckstr. 1,
D-64291 Darmstadt, Germany}
\affiliation{Moscow  Institute for Physics and Engineering, Kashirskoe
sh. 31, RU-115409 Moscow, Russia}
\begin{abstract}

A QCD-motivated dynamical-quasiparticle model with parameters adjusted to
reproduce the lattice-QCD equation of state is extrapolated from
region of high temperatures and moderate baryonic densities to the
domain of high baryonic densities and zero temperature. The resulting
equation of state  matched with realistic
hadronic equations of state predicts a phase transition into the quark phase
at higher densities than those reachable in neutron
star interiors. This excludes the possibility of the existence of hybrid
(hadron-quark) stars. Pure quark stars are  possible and
have low masses, small radii and very high central densities.
Similar results are obtained for a simple bag model with massive
quarks, fitted to reproduce the same lattice results. Self-bound
quark matter is also excluded within these models. Uncertainties in the
present extrapolation are discussed. Comparison with standard bag
models is made.
\end{abstract}
%\date{\today}

\maketitle

\section{Introduction}

Nowadays it is commonly accepted that the quark-gluon phase of the
matter can be formed in the course of heavy-ion collisions at
ultrarelativistic energies. There are indirect signatures that
this state has been produced at Relativistic Heavy-Ion Collider
(RHIC)~\cite{GM04}. On the other hand, it was suggested long
ago~\cite{Itoh} that quark matter may exist either in the interior
of some  stars or as a new family of pure quark stars. Then it was
realized that some rather massive neutron stars may have a quark
core (so called hybrid stars), cf.
Refs \cite{G97,W04}
and
references therein. Recently the idea of quark pairing with high
values of the gap (up to $\Delta \sim 100~$MeV) attracted much
attention, cf. Ref. \cite{RW}. Observable signals  of quark
matter, presumably existing in interiors of hybrid and pure quark
stars, are under extensive discussion. The most promising of these
signals seem to be an abnormal mass-radius relation~\cite{bomb},
$r$-mode spin-down~\cite{M00}, and a specific steller cooling
properties~\cite{bkv}.

Although there is no {\it a priori} obstacle to existence of the
quark matter in above the mentioned forms, the actual theoretical predictions
are very uncertain. They hinge on the lack of understanding of the confinement
mechanism in QCD. Therefore, parameters of theoretical quark models cannot
be anchored to experimental observables.
Most of the treatments are based on either the standard or slightly
modified MIT bag model, or the Nambu--Jona-Lasino (NJL) model,
cf. Ref. \cite{B04} and references therein.
In terms of a bag model, the main uncertainty is associated with the
value of the vacuum pressure, i.e. the bag constant $B_{\rm eff}$.
The steller properties are sensitive to this. A large
value of $B_{\rm eff}$  excludes the existence of hybrid stars. With a
decrease of $B_{\rm eff}$, first the most massive neutron stars acquire
a quark core, becoming hybrid stars.
For a small $B_{\rm eff}$, strange quark matter becomes
absolutely stable, i.e. more stable than
$^{56}\mbox{Fe}$~\cite{W84,AFO}. Then pure quark stars
(strangelets)
of an arbitrary size may exist, being hold together  by strong
rather than gravitational forces.
Physical implications of the existence of  a self-bound strange
quark matter are reviewed in Ref.~\cite{selfbound}.

A straightforward calculation of the properties of strongly
interacting matter and, in particular its equation of state (EoS),
is possible on the lattice~\cite{latEoS0}. This technique
is progressing rapidly. So far, the main body of lattice results
concern the case of zero baryon chemical potential ($\mu=0$) and
finite temperatures. This case is just opposite to that of
cold baryon matter present in the quark stars. Recently new
lattice predictions on the EoS  at finite, but small, baryon chemical
potentials became  available \cite{latEoS,Allton,Gavai}.
Nevertheless, they are still far from the
astrophysically relevant domain. Therefore, a theoretical modeling
is needed for their extrapolation.

The interpretation of these lattice results within phenomenological
quasiparticle models,
i.e. in terms of effectively massive quarks and gluons with a simple
interaction, turned out to be very successful both at zero chemical
potential~\cite{Gorenstein95,Greiner,Levai,Pesh96,Pesh00,Pesh00b,Weise01,Rebhan03,Szabo03,Weise04,BKS,I04}
and at small finite ones~\cite{Szabo03,BKS,Weise04,I04}.
With few phenomenological parameters it was possible to
reasonably reproduce all lattice thermodynamic quantities.
%% at zero baryon chemical potential.
Effectively the results can be interpreted in terms of a
modified MIT bag model for massive quarks with  density and temperature
dependent bag constant and  masses. These models can be used to
extrapolate the lattice EoS to the domain of cold baryon matter.
This has been done by Peshier, K\"ampfer and Soff~\cite{Pesh00} for the
thermodynamic quasiparticle model of  Ref.~\cite{Pesh96}. They
considered the possibility of the existence of quark
stars  and found a mass-radius relation for them. They concluded that
quark stars with a mass
larger than $1M_{\odot}$ may only exist for values of the effective bag
constant $B^{1/4}\leq 180-200$~MeV, whereas their model produced a
higher value. The possible existence of hybrid stars was not considered in that
work. It is important to mention that the quasiparticle model of
Ref.~\cite{Pesh00} was fitted only to the $\mu=0$ lattice data
available at that time and then it was extrapolated to finite values
of the baryon chemical potential.
The later fit of the same model to recent lattice data at finite baryon
chemical potentials~\cite{latEoS} resulted in somewhat different
parameters of this quasiparticle model~\cite{Szabo03}.
The same quasiparticle model of Ref.~\cite{Pesh96} was used in
Ref.~\cite{Rebhan03} in order to extrapolate the HTL calculations to
the cold-baryon-matter domain. Conclusions very similar to those in
Ref.~\cite{Pesh00} were derived concerning the properties of quark
stars.

In this paper we would like to make ``lattice QCD motivated''
predictions on the possible existence of hybrid and quark stars, as
well as their properties. To do this, we extrapolate the existing
lattice results to the region of cold electroneutral baryon matter
relying on the phenomenological QCD motivated
dynamical-quasiparticle (DQ) model recently proposed in Ref.
\cite{I04}. In equilibrium, at high temperatures the DQ model
complies with the quasiparticle picture of the hard thermal loop
approach~\cite{Pisarski}, whereas at lower temperatures it
simulates the confinement of the QCD. The latter is manifested by
the fact that the solution to the model equations simply does not
exist below certain combination of the temperature and the
chemical potential. Two sets of parameters of this model (below
denoted as ``1-loop'' and ``2-loop'') were fitted to reproduce
lattice data at finite baryon chemical potentials~\cite{latEoS}.
Extrapolation of the lattice data to the case of the cold baryon
matter is essentially simpler in this model than in that of
Ref.~\cite{Pesh00}, where an implicit thermodynamic-consistency
equation should be numerically solved. In fact, this advantage
follows from an explicitly thermodynamically consistent formulation
in terms of dynamic rather than thermodynamic variables.

In Sect.~\ref{Models} we start with brief recapitulation of the DQ
model. To have a reference point for the DQ model,
we also consider three versions of the MIT bag model.
Two ``light-bag''
models (``light-bag-155'' and ``light-bag-200'') with conventional
current quark masses ($m_{u0}=5$~MeV, $m_{d0}=7$~MeV and
$m_{s0}=150$~MeV), zero gluon mass ($m_g=0$) and
different bag constants ($B^{1/4}=$ 155~MeV and 200~MeV) covering a broad range
of EoS's  usually applied to the treatment of hybrid and quark
stars~\cite{HHJ,ABPR}.
We also consider a ``heavy-bag'' model, the quark/gluon masses and bag
constant of which are fitted to reasonably reproduce the
lattice results of Ref.~\cite{latEoS}:  $m_u=m_d=330$~MeV,
$m_s =450$~MeV, $m_g =600$~MeV and $B^{1/4}=183$~MeV.
In Sect.~\ref{Comparison} we confront the predictions of all these
models to the (2+1)-flavor lattice data~\cite{latEoS} at finite
chemical potentials. In Sect.~\ref{Stars} we derive the
predictions of all the above models regarding the possible existence of hybrid and
quark stars and discuss their properties.

\section{Models}\label{Models}

\subsection{Dynamical-Quasiparticle Model}

The effective Lagrangian of the DQ model \cite{I04} treats
transverse gluons $\phi_a$ and quarks $\psi_{fc}$ of $N_f=3$
flavors and $N_c=3$ colors. For this model it is essential
  that $N_f=N_c$. These constituents interact via mean fields
${\vec\zeta}= \{\zeta_u,\zeta_d,\zeta_s\}$:
\begin{eqnarray}
\label{qg-lagr}
{\cal L} &=& \frac{1}{2}\sum_{a=1}^{N_g}
\left[(\partial_\mu \phi_a)^2 - m_g^2({\vec\zeta}) \phi_a^2\right]
\cr &+&
\sum_{c=1}^{N_c}\sum_{f=1}^{N_f}
\bar{\psi}_{fc} [i\gamma_\mu \partial^\mu
- m_{f}({\vec\zeta})]\psi_{fc}
-B(\chi),
%-\frac{1}{2}U(\eta,{\vec\xi}),
\end{eqnarray}
where $N_g=2(N_c^2-1)$ is the number of transverse gluons,
\begin{eqnarray}
\label{chi}
\chi^4 = \sum_{f=1}^{N_f} \zeta_{f}^4,
\end{eqnarray}
\begin{eqnarray}
\label{m_g-qg-s}
m_g^2  &=&
\frac{2}{N_g} \sum_{f=1}^{N_f} \zeta_{f}^2 g^2(\chi),
\\
m_{f}^2-m_{f0}^2 &=&
\frac{1}{2N_c} \zeta_{f}^2 g^2(\chi).
\label{m_q-qg-s}
\end{eqnarray}
Here $m_{f0}$ is the current $f$-quark mass, whereas $m_g$ and $m_f$
are effective masses of gluons ($g$) and quarks of flavor  $f$,
respectively. The masses and the coupling constant $g^2(\chi)$ depend
on mean fields ${\vec\zeta}$.
\begin{eqnarray}
B(\chi)
&=& B_C-
\frac{1}{N_g}\left[\chi^4 g^2(\chi)-\chi_C^4 g^2(\chi_C)\right]
\nonumber\\
&+&
\frac{2}{N_g}\int_{\chi_C}^{\chi} d\chi_1 \chi_1^3g^2(\chi_1)\,,
\label{U_g}
\end{eqnarray}
is the potential of the mean-field self-interaction having a
meaning of an effective bag parameter. Here $B_C$ and $\chi_C$ are
parameters of the model.

The equations of motion for the mean fields read
\begin{eqnarray}
-\frac{\partial{\cal L}}{\partial\zeta_i^2}&=&
\frac{\partial B}{\partial\zeta_i^2}+
\frac{1}{2}\frac{\partial m_g^2}{\partial\zeta_i^2} \eta^2
%\nonumber\\
%&+&
+
\frac{1}{2}\sum_{f=1}^{N_f}
%\frac{\partial (m_f^2-m_{f0}^2)}{\partial\zeta_i^2}
\frac{\partial m_f^2}{\partial\zeta_i^2}
\xi_f^2=0\,,
\label{qg-gap-s}
\end{eqnarray}
where
\begin{eqnarray}
\label{g-rho1}
\eta^2 &\equiv&\sum_{a=1}^{N_g}\left\langle \phi_a^2\right\rangle=\frac{N_g}{2\pi^2}
\int_0^\infty \frac{k^2\;dk}{(k^2+m_g^2)^{1/2}} f_g (k),
\\
\xi_{f}^2 &\equiv& \sum_{c=1}^{N_c}
\frac{\left\langle \bar{\psi}_{fc}\psi_{fc}\right\rangle}{m_{f}}
\nonumber\\ &=&
\frac{N_c}{\pi^2}
\int_0^\infty \frac{k^2\;dk}{(k^2+m_{f}^2)^{1/2}}
[f_{q,f} (k)+\bar{f}_{q,f} (k)],
\label{q-rho1}
\end{eqnarray}
are scalar densities of gluons and quarks (divided by the mass),
respectively, and $f_g (k)$, $f_{q,f} (k)$ and $\bar{f}_{q,f} (k)$ are
occupation numbers (distribution
functions) of gluons, quarks and antiquarks.
In the case of equilibrium that we consider here, these are
\begin{eqnarray}
\label{fg}
f_g (k)&=&\frac{1}{\exp[(k^2+m_g^2)^{1/2}/T]-1},
\\
\label{fq}
f_{q,f} (k)&=&\frac{1}{\exp\{[(k^2+m_f^2)^{1/2}-\mu_f]/T\}+1},
\\
\label{faq}
\bar{f}_{q,f} (k)&=&\frac{1}{\exp\{[(k^2+m_f^2)^{1/2}+\mu_f]/T\}+1},
\end{eqnarray}
where  $T$ is the temperature, and $\mu_{f}$ is the  $f$-flavor
quark chemical potential. In the general case, all $\mu_{f}$ may be different.

The solution of above equations of motion (\ref{qg-gap-s}) with
respect to the mean fields is
\begin{eqnarray}
\label{zeta}
\zeta_{f}^2 =\eta^2 + \frac{N_g}{4N_c}\xi_{f}^2.
\end{eqnarray}
Here we heavily rely on the condition $N_c=N_f$.
It is easy to verify~\cite{I04} that with this solution the effective
masses of
gluons and quarks, cf. Eqs (\ref{m_g-qg-s}) and (\ref{m_q-qg-s}),
reproduce the hard-thermal-loop results~\cite{Pisarski} in the
high-temperature limit, provided the coupling constant $g^2(\chi)$ is
appropriately defined. In fact, this property was the main
requirement in construction of this DQ model \cite{I04}.
The appropriate choice of the coupling constant~\cite{I04} is the
following
\begin{eqnarray}
\label{g2_eff}
g^2(\chi)  =
\frac{16\pi^2}{ \beta_0\ln[(\chi^2+\chi_0^2)/\chi_C^2]}
f(\chi)\,,
\end{eqnarray}
where $\beta_0 = \frac{1}{3}(11 N_c - 2 N_f^{\scr{eff}})$.
$N_f^{\scr{eff}}$ is the
effective number of quark flavors at the energy scale $\chi$, which
may be   $N_f^{\scr{eff}}<N_f$ \footnote{Let us remind that $N_f=N_c$.}.
The parameter $\chi_C^2$ is introduced in Eq.~(\ref{U_g})
and $\chi_0^2$ is another phenomenological parameter of the model.
An auxiliary function $f(\chi)$, satisfying the condition
$f(\chi\to\infty)\to 1$, helps us to choose between the 1-loop
and 2-loop perturbative limits of the coupling constant \cite{Kapusta,Yndurain}.

The ``1-loop'' version of the model
\begin{eqnarray}
\label{1-loop}
f_{\rm 1-loop}\equiv 1
\end{eqnarray}
 overestimates by approximately 10\%
the lattice data (see Ref. \cite{I04} and discussion in
Sect. \ref{Comparison}). Therefore, the ``2-loop'' version was also
considered. The latter was chosen in the form
\begin{eqnarray}
\label{2-loop}
f_{\scr{2-loop}}(\chi)=
1 + \arctan \left[\frac{\beta_1}{8\pi^2\beta_0}
g^2(\chi)\ln\frac{g^2(\chi)}{\lambda}\right]
\end{eqnarray}
with
$$
\beta_1 =
\frac{1}{6}(34 N_c^2 - 13 N_c N_f^{\scr{eff}} + 3 N_f^{\scr{eff}}/N_c),
\,\,
\lambda =  0.001\frac{16\pi^2}{\beta_0}.$$
With this
$f_{\scr{2-loop}}$, the coupling constant meets the 2-loop perturbative
limit at $\chi\to\infty$.
The additional $g^2 \ln (0.001)$ term is sub-leading as compared to
$g^2(\chi)\ln g^2(\chi)$ and thus does not prevent agreement with the
2-loop approximation for the coupling constant. In fact, the function
$f_{\scr{2-loop}}$ is an ``exotic'' representation of a constant,
since in the range, covered by lattice simulations,
$f_{\scr{2-loop}}(\chi)\simeq$ 2.6 with good
accuracy. Precisely this enhancement of the coupling constant is
required to fit the actual overall normalization of the lattice
results.

As we have  mentioned the present model
simulates the confinement of quarks and gluons.
At some value of $\chi$ the argument of
$\ln[(\chi^2+\chi_0^2)/\chi_C^2]$ in Eq. (\ref{g2_eff}) becomes very
close to 1, and hence $g^2\to\infty$. Due to that there are no solutions to
the above equations below certain values of temperature and chemical
potentials. This can be interpreted as a kind of confinement.

\begin{table*}
\begin{ruledtabular}
\begin{tabular}{cccccccc}
 Version& $T_C$\footnote{The critical temperature {\small $T_C$} is chosen on the
condition of the best reproduction of lattice predictions \cite{latEoS} by
the ``heavy-bag'' model. For the ``light-bag'' models it was simply
kept the same.}, MeV & $m_u$, MeV & $m_d$, MeV & $m_s$, MeV &
$m_g$, MeV & $B^{1/4}$, MeV & c
\\ \hline
light-bag-155& 167 &   5 &   7  & 150 &   0 & 155 & $0\div 0.3$\\
light-bag-200& 167 &   5 &   7  & 150 &   0 & 200 &  0         \\
heavy-bag    & 167 & 330 & 330  & 450 & 600 & 183 & $0\div 0.3$\\
\end{tabular}
\caption{Parameters of bag models}
\label{tab:bagpar}
\end{ruledtabular}
\end{table*}

To summarize, the procedure of solving the model equations is as
follows. First, we define all the free parameters of the model
($\chi_C$, $\chi_0$, $B_C$), including the auxiliary function
$f(\chi)$. Given the temperature $T$ and the set of chemical
potentials $\mu_{f}$, the set of equations
(\ref{chi})--(\ref{m_q-qg-s}) and
(\ref{g-rho1})--(\ref{g2_eff}) should be solved.
As a result of this solution, we obtain effective quark and gluon
masses and the value of $\chi$ variable, which is required for
calculation of $B(\chi)$, cf. Eq. (\ref{U_g}).
When all the quantities are defined, we can calculate the energy
density $\varepsilon(T,\{\mu_f\})$, pressure $P(T,\{\mu_f\})$, and
baryon density $n(T,\{\mu_f\})$ as follows
\begin{widetext}
\begin{eqnarray}
\label{E_qg-s}
\hspace*{-9mm}
\varepsilon(T,\{\mu_f\})&=&\left(1-\frac{4}{5}c\right) \frac{N_g}{2\pi^2}
\int_0^\infty k^2\;dk\;(k^2+m_g^2)^{1/2}\; f_g (k)
\cr &+&
\left(1-c\right) \sum_{f=1}^{N_f}
\frac{N_c}{\pi^2}
\int_0^\infty k^2\;dk\;(k^2+m_{f}^2)^{1/2}\; [f_{q,f}(k)+\bar{f}_{q,f}(k)]
+B(\chi)+\delta B_{\Delta},
\\
\label{P_qg-s}
\hspace*{-9mm}
P(T,\{\mu_f\})&=&\left(1-\frac{4}{5}c\right)\frac{N_g}{6\pi^2}
\int_0^\infty \frac{k^4\;dk}{(k^2+m_g^2)^{1/2}} f_g (k)
\cr &+&
\left(1-c\right) \sum_{f=1}^{N_f}
\frac{N_c}{3\pi^2}
\int_0^\infty \frac{k^4\;dk}{(k^2+m_{f}^2)^{1/2}} [f_{q,f}(k)+\bar{f}_{q,f}(k)]
-B(\chi)-\delta B_{\Delta},
\\
\label{n_qg-s}
\hspace*{-9mm}
n(T,\{\mu_f\})&=&\frac{1}{3}\left(1-c\right) \sum_{f=1}^{N_f}
\frac{N_c}{\pi^2}
\int_0^\infty k^2\;dk\; [f_{q,f}(k)-\bar{f}_{q,f}(k)].
\end{eqnarray}
\end{widetext}
The gluon and antiquark terms vanish at zero temperature. Here we have introduced
factors $\left(1-\frac{4}{5}c\right)$ and $\left(1-c\right)$ to make
provision for the bag model (see the next Subsect.). For the DQ
model, $c$ is identically  zero, $c_{\scr{DQ-model}}\equiv 0$.
%since the
%corresponding corrections are included in a self-consistent fashion.

In Eqs (\ref{E_qg-s}) and (\ref{P_qg-s}) we have also made
provision for a pairing
contribution $\delta B_{\Delta}$, which may be important at low
temperatures. The respective term can be directly added to the
Lagrangian. It preserves the above mean-field solution, provided it
is independent of these mean fields. For temperature higher than a certain temperature
$T_{c}^{\rm CSC}$, this contribution vanishes, $\delta B_{\Delta}=0$.
For $T<T_{c}^{\rm CSC}$ the quarks can be paired, forming a color
superconductor. This gives rise to $\delta B_{\Delta}$
that depends on the color
superconducting gap. The form of this term, $\delta B_{\Delta}$, depends
on the pairing channel. There exists a variety of possible phases. We will
present the results for  two phases usually discussed.
One expects that the phase that can be realized at
moderate densities
%%sufficiently small values of $\mu_s$, i.e. when there is no $s$ quark  Fermi
%%sea,
is
the two-color-superconducting (2SC) phase.  At larger densities the
most symmetric
color-flavor-locked (CFL) phase may be formed (for $\Delta >3m_s^2 /(2\mu )$). In these cases we
have \cite{Shovk04}
\begin{eqnarray}
\delta B_{\Delta}\simeq -\frac{\nu \Delta^2\mu^2}{\pi^2}
\quad \mbox{at} \quad T<T_{c}^{\rm CSC},
\end{eqnarray}
 where $\Delta (T)$ is the gap,
$\nu =1/9$ for 2SC phase and $\nu =1/3$ for CFL phase, and $\mu$ is the
baryon chemical potential.

Note that thermodynamic consistency is automatically fulfilled in
our scheme, since we proceed from a proper Lagrangian formulation.
The presented model is of pure mean-field nature. In particular, it
disregards fluctuations that may be important in
the vicinity of both the deconfinement and the superconducting phase
transitions \cite{Vfl04}.
However, these fluctuations
are usually disregarded in other models applied to both
high and low temperatures as well. These effects require further study.

\subsection{Bag models}

The last formulas of the previous subsection, i.e. Eqs
(\ref{E_qg-s})--(\ref{n_qg-s}), are directly applicable to the bag
model with the only simplification that all masses $m_u,m_d,m_s,m_g$
and the bag constant $B$ are really constants, i.e. are independent of
any mean fields, temperature and chemical potentials. Often
perturbative corrections $\propto g^2$
to the kinetic terms of both quarks \cite{Jaffe} and gluons
\cite{Zhai} given by the factor $c$ (which is
$2\alpha_s^2 /\pi$ in case of massless quarks and gluons) are also taken into account, see Eqs
(\ref{E_qg-s})--(\ref{n_qg-s}). Since perturbation theory
is not applicable to the whole region of temperatures and
chemical potentials of our interest,
we will not use an explicit expression for this correction but rather  vary it
within a reasonable range, $c=0\div 0.3$, following Ref. \cite{ABPR}.

Below we consider three versions of the bag model, the parameters are listed in Table~\ref{tab:bagpar}.
The two first versions, labeled ``light-bag-155'' (similar to the model
used in Ref. \cite{ABPR}) and
``light-bag-200'' (similar to the model used in Ref. \cite{HHJ}),  represent a range of ``conventional''
bag models
used in astrophysical applications. As we will show below these models
cannot reproduce lattice results.
The ``heavy-bag'', as well as  DQ based models,  are fitted
to reasonably reproduce lattice data of Ref. \cite{latEoS}, see next
section. This fit required heavy quarks and gluons to be introduced.

%------------------------------------------------------------------------
\begin{figure*}
\includegraphics[height=60mm,clip]{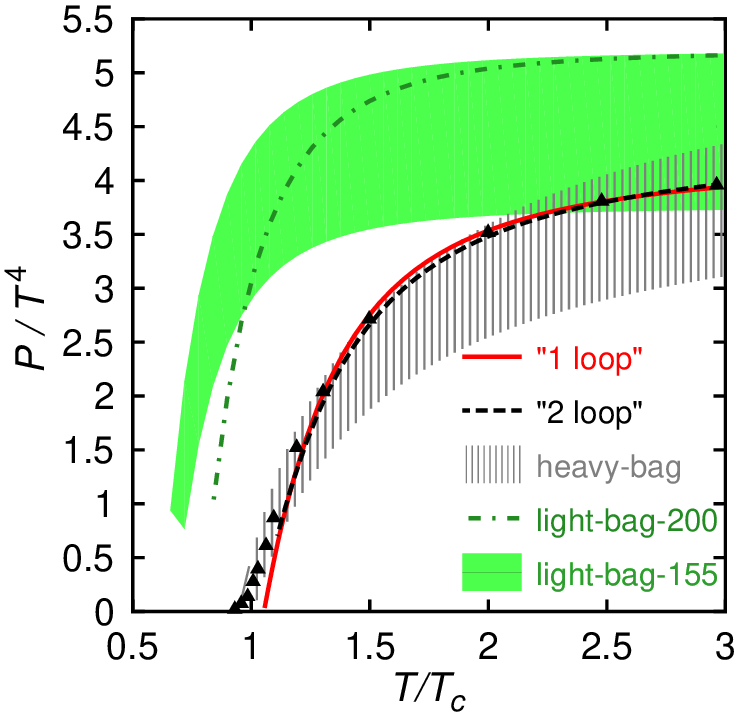}\quad
\includegraphics[height=60mm,clip]{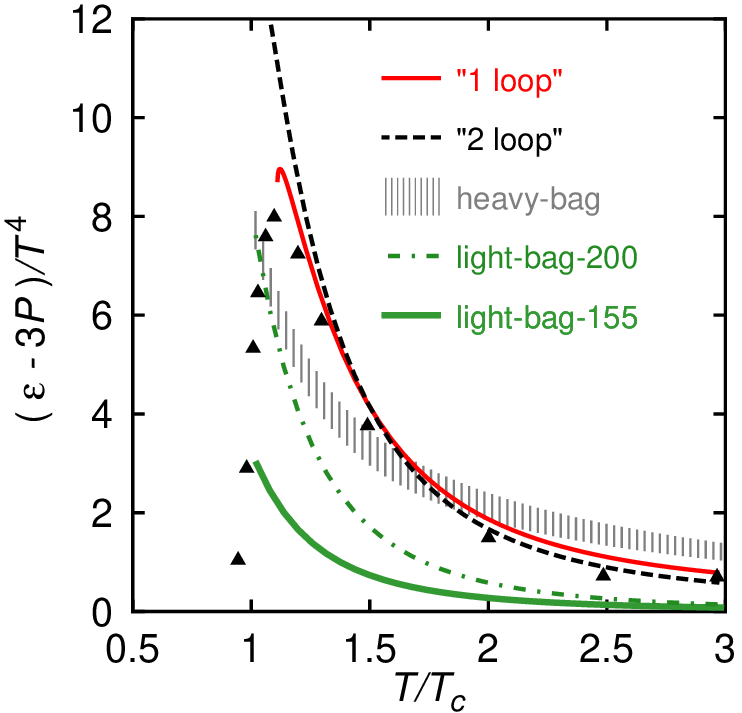}
\caption{(Color online)
Pressure (left panel) and the interaction measure, $\varepsilon-3P$,
  (right panel) scaled by $T^4$ as functions of $T/T_C$ at
  zero baryon chemical potential $\mu=0$.
The solid line corresponds to ``1-loop'' version, whereas the
  dashed line, to the ``2-loop'' one of the DQ model.
The vertically hatched band is the ``heavy-bag'' fit
with $c=0$ and $c=0.3$ for the top and bottom
  boundary lines, respectively.
The grey band (for $P$) and the grey line (for $\varepsilon-3P$)
  represent prediction of the ``light-bag-155'' model.
The dash-dotted line corresponds to the ``light-bag-200'' model.
Lattice data, triangles and dots, are from Ref. \cite{latEoS}.
} \label{p0_fig}
\end{figure*}

\section{Comparison with Lattice Results}\label{Comparison}

In this section we demonstrate the quality of  the fit of the above
described models to the recent (2+1)-flavor lattice predictions at various
chemical potentials \cite{latEoS}.

We again start with a brief recapitulation of Ref. \cite{I04},
where this comparison was done for the DQ model. To be consistent
with the lattice data, current quark masses $m_{u0}=m_{d0}=$ 65
MeV and $m_{s0}=$ 135 MeV were adopted, as in these lattice
calculations. It was found that the actual results of our
quasiparticle model are quite insensitive to variation of $m_{f0}$
from above lattice values to the ``physical'' ones
$m_{u0}=m_{d0}=$ 7 MeV and $m_{s0}=$ 150 MeV. This
insensitivity is to the extent that predictions of the DQ model
with lattice and ``physical'' quark masses are hardly
distinguishable in the scale of figures presented below. In fact,
this is not surprising, since the 80$\div$90\% of the effective
quark masses (\ref{m_q-qg-s}) result from interaction with the
mean field $B(\chi)$.  The model involves several
phenomenological parameters: the parameter $B_C$, cf. (\ref{U_g}),
the ``QCD scale'' $\chi_C$, and  an auxiliary function $f(\chi)$,
cf. (\ref{g2_eff}). Another parameter $\chi_0^2$ should be taken
small $\chi_0^2\ll\chi_C^2$. It shifts the lower limit of
integration in the expression for $B(\chi)$, cf. Eq. (\ref{U_g}),
from the singular point of the coupling constant,  cf. Eq.
(\ref{g2_eff}), and hence regularizes the calculation of
$B(\chi)$. Therefore, it is closely related to the parameter
$B_C$, which is an integration constant in the same expression. A
change of $\chi_0^2$ implies the corresponding change of $B_C$. In
all further calculations we take $\chi_0^2= 0.01 \chi_C^2$, and
the values of $B_C$ stated below correspond only to this choice.

An implicit parameter
of the model is the critical temperature $T_C$, i.e. the temperature
at which the deconfinement phase transition occurs at $\mu=$ 0.
It would not be reasonable to
identify this temperature with that of the end point of the
solution discussed above. The reason is that the phase transition at
$\mu=$ 0 in the case of (2+1) flavors is of the cross-over type, as
found in lattice calculations. This implies that a strong
interplay between quark--gluon and hadronic degrees of freedom occurs
near $T_C$, which actually determines the $T_C$ value itself.
%%Moreover, in Ref. \cite{Redlich} it was
%%argued that below $T_C$ the thermodynamic quantities are quantitatively
%%well described by a model of the resonance hadronic gas.
Since hadronic degrees of freedom are completely disregarded in the
model, it cannot properly determine the $T_C$ value.
Therefore, the value of
$T_C$  was varied from the determined end-point temperature to slightly below
in order to achieve the best fit of the lattice results.
The fitted value $T_C=$ 195 MeV is slightly above its lattice
counterpart 175 MeV.

As demonstrated in Ref. \cite{I04}, the ``1-loop'' version of the DQ
 model, cf. Eq. (\ref{1-loop}), does not reproduce the
overall normalization of the lattice pressure.
For the best fit of the lattice data the overall normalization factor
($N_{\rm norm}$) was chosen equal 0.9 for the
``1-loop'' calculations.  The value $N_{\rm norm}$ is indeed
 somewhat uncertain.
Lattice calculations were done
on lattices with $N_t=$ 4 temporal extension \cite{latEoS}. To
transform the ``raw'' lattice data into physical values, i.e. to extrapolate
to the continuum case of $N_t\to\infty$, the raw data are multiplied
by {\em ``the dominant $T\to\infty$ correction factors between the
  $N_t=$ 4 and continuum case''}, $c_p=$ 0.518 and $c_\mu=$ 0.446,
\cite{latEoS}. These factors are determined as  ratios of the
Stefan--Boltzmann pressure at $\mu=$ 0 ($c_p$) and the $\mu$-dependent
part of the Stefan--Boltzmann pressure ($c_\mu$) to the corresponding
values on the $N_t=$ 4 lattice \cite{latEoS}. In view of this
uncertainty, it is reasonable to apply an additional overall
  normalization factor $N_{\rm norm}$ to the same quantities calculated within
DQ model \footnote{It would be more reasonable to apply this
  additional normalization factor to the lattice data. However, we do
  not want to distort the ``experimental'' results.}. In order to keep
the number of fitting parameters as small as possible, a single
normalization factor $N_{\rm norm}$ was applied instead of two
 different ones, $c_p$  and $c_\mu$, as in the lattice calculations.

\begin{table}
\begin{ruledtabular}
  \begin{tabular}{cccccccc}
Version& $N_f^{\scr{eff}}$ & $T_C$, MeV & $\chi_C$, MeV &
$B_C/\chi_C^4$\footnote{ These $B_C$ values correspond to the choice
$\chi_0^2=0.01 \chi_C^2$.}
& $f$-factor\footnote{ This is the effective value of the auxiliary function
  $f(\chi)$ in the temperature range under investigation, i.e. from
  $T_C$ to $3T_C$.}  & $N_{\rm norm}$\\ \hline
 ``1-loop''& 3          & 195 & 141.3 & -97.5  & 1 & 0.9  \\
 ``2-loop''& 3          & 195 & 119.6 & -267.5 & 2.6 & 1  \\
 \end{tabular}
\end{ruledtabular}
\caption{Best fits of DQ-model parameters to lattice data~\cite{latEoS}}
\label{tab:dqpar}
\end{table}

As for the ``2-loop'' version,
 the auxiliary function (\ref{2-loop}) was chosen to reproduce the
 lattice data normalization.
In fact, the function $f_{\scr{2-loop}}$
is an ``exotic'' representation of a constant,
since in the temperature range under consideration, from $T_C$ to
$3T_C$, $f_{\scr{2-loop}}(\chi)\simeq$ 2.6 with good accuracy.
In this respect, any function $f$, providing us with the additional
factor 2.6 in the temperature range from $T_C$ to $3T_C$ and
respecting the proper perturbative limit of the coupling constant at
 very high temperatures, is just as suitable for this fit.
The two sets of parameters of the DQ  model are summarized in Table~\ref{tab:dqpar}.

\begin{figure*}
\includegraphics[height=60mm,clip]{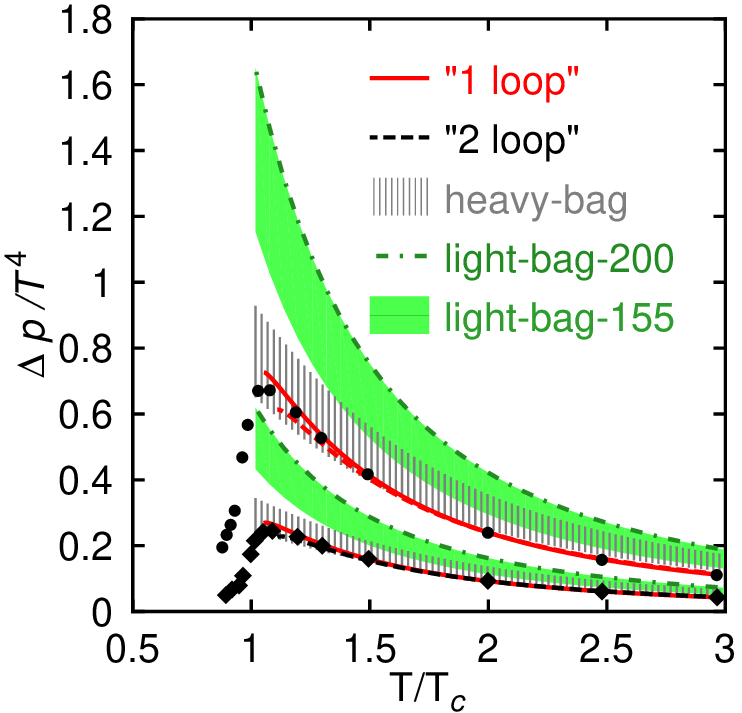}
\quad
\includegraphics[height=60mm,clip]{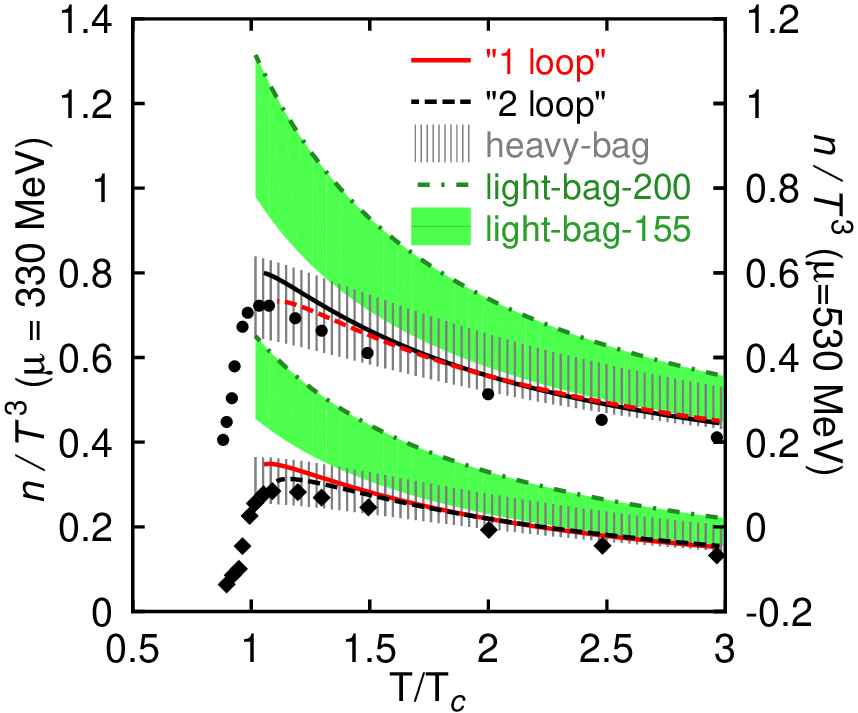}
\caption{(Color online)
$\Delta P = P(T,\mu)-P(T,\mu=0)$ (left panel) and the
 baryon density $n$ (right panel)
scaled by $T^4$ and $T^3$ as functions of $T/T_c$ at nonzero baryon chemical
potentials $\mu =$ 330 and
530 MeV (from bottom to top). Note different scales (left and right, respectively)
for $\mu =$330 and 530 MeV in the right panel.
Notation is the same as in Fig.~\ref{p0_fig}.
} \label{dp}
\end{figure*}
%------------------------------------------------------------------------

%------------------------------------------------------------------------
\begin{figure*}
\includegraphics[height=64mm,clip]{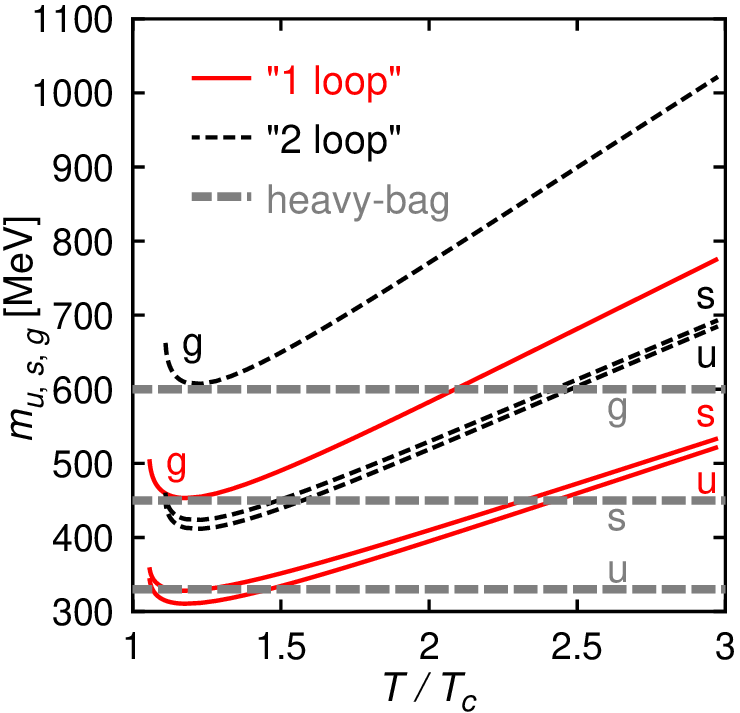}
\quad
\includegraphics[height=64mm,clip]{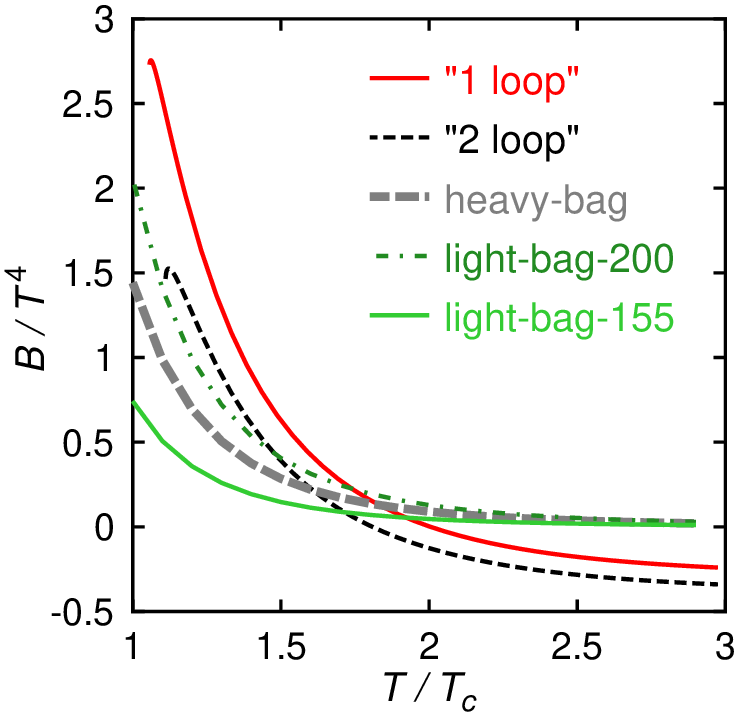}
\caption{(Color online)
Quark and gluon effective masses (left panel) and bag
  parameters scaled by $T^4$ (right panel)
  for the DQ model, cf. Eq. (\ref{U_g}), and bag models
 as functions of $T/T_c$ at zero
  baryon chemical potential $\mu =0$.
Notation is the same as in Fig.~\ref{p0_fig}.
} \label{u_t}
\end{figure*}
%------------------------------------------------------------------------

The comparison with (2+1)-flavor lattice results \cite{latEoS} of the
above two versions of the DQ model, as well as the various bag models of Table~\ref{tab:bagpar},
is presented in Figs.~\ref{p0_fig} and \ref{dp}. The baryon chemical
potential $\mu$ is defined as follows: $\mu_u=\mu_d=\mu/3$ and
$\mu_s=0$. {\em{Both versions of
the DQ model perfectly fit the lattice data above $T_C$.}}
This fact demonstrates
that the way to fit these data even within the DQ model is not unique.
Moreover, as it was demonstrated in Ref. \cite{I04}, even 2-flavor
lattice pressure \cite{Allton} is well reproduced above $T_C$ by the
DQ model with practically the same parameters.

For two of the bag models we have taken into account the uncertainty
associated with the perturbative correction $c$. Following
Ref. \cite{ABPR} we varied $c$ from 0 to 0.3, see Table~\ref{tab:bagpar}. Therefore,
the results for bag models (``light-bag-155'' and ``heavy-bag'')
are displayed, as a rule, by bands with
$c=0$ and $c=0.3$ for the top and bottom boundary lines, respectively.
In the case of the ``interaction  measure'' $\epsilon -3P$ and
light quarks, the kinetic terms cancel with good accuracy and the band
shrinks to a line. For the ``light-bag-200'' model, we do
not display uncertainties associated with $c$ in order not to overload the
figure
and since the general picture seems clear without this additional variation:
{\em{the ``light-bag'' models do not follow lattice predictions.}}
%%The conventional ``light-bag'' models fail to reproduce the lattice data.
 We checked that these fits to the pressure and ``interaction
 measure'' $\epsilon -3P$ could be improved
at the price of a significant increase of the bag constant $B$. However,
reproduction of these lattice quantities still remains rather poor
even with this improvement. Moreover, the remaining quantities, $\Delta
P$ and $n_B$, prove to be unchanged, since they are independent of  $B$.
Within the above mentioned $c$-uncertainty,
{\em{the ``heavy-bag'' model reasonably well covers the lattice data,
however, at expense of introducing large quark and gluon constituent masses.}}

To reveal similarities between the DQ and bag models, in
Fig.~\ref{u_t} we compare quark/gluon masses and bag parameters from
these two kinds of models. Note that these are the only quantities that
matter in the calculations of thermodynamic functions such as those
of Eqs (\ref{E_qg-s})--(\ref{n_qg-s}). The $u$ and $d$ quark masses of
the DQ model are indistinguishable within the scale of this figure.
Quark and gluon masses,
%%These quantities
required for the reproduction of the lattice results,
%%of the same order
are high ($\gsim 300~$MeV) in both the DQ
model and the ``heavy-bag'' model. Thus, {\em{the failure of the ``light-bag''
models to reproduce lattice data can be associated with the light quark and gluon masses used in the
calculation.}} Note that in the DQ
model, $B$ becomes negative at high temperatures. This means that the
interaction changes from repulsion to attraction. This behavior of the bag
parameter is common for models reproducing  lattice data, cf. Refs \cite{Pesh96,Weise01}.

As argued in Ref. \cite{Redlich},
the lattice data below $T_C$ are
well reproduced by the resonance hadronic gas model.
However to comply with the
lattice results, hadron masses in this resonance hadronic gas should be
enhanced. In particular, the pion mass was taken
$m_{\pi}\simeq 770$~MeV, as resulted from lattice
calculations. This fact indicates that the EoS in the hadronic sector is
not reliably predicted by the lattice calculations.
%, even for high temperatures and small baryon chemical potentials.
The analysis of Ref. \cite{V04} shows that strong interaction effects
might be crucially important in the description of the hadron system at high
temperatures and small baryon chemical potentials. Also we cannot use the
resonance hadronic gas model at the low temperatures of present interest due to
the lack of interactions incorporated in this model.
This is why we avoid fitting our realistic hadronic EoS's, discussed
in the next section, to the lattice data.

\section{EoS at $T=0$: hybrid and quark stars}\label{Stars}

Possible existence of  compact stars consisting either in part
(hybrid stars)  or completely (quark stars) of a deconfined
quark matter is one of the vividly discussed aspects of
modern astrophysics, see recent
reviews~\cite{Bomb02,Xu02,Schaffner04}. Typically, for the
description of a deconfined phase a certain version of the MIT bag
model is used, see Ref.~\cite{Drago01} for a survey of
parameterizations. In this section we explore a possibility for
hybrid and quark stars, proceeding from the DQ and heavy-bag models
for the EoS of deconfined matter.

\subsection{Cold star matter}

For matter in $\beta$-equilibrium at zero temperature
several conditions should be met.
Chemical potentials of quarks and leptons satisfy equilibrium
conditions with respect to reactions
$u+e\leftrightarrow s$, $d\leftrightarrow s$, $\mu\leftrightarrow e$:
\begin{eqnarray}
\mu_u +\mu_e =\mu_s \,,\quad \mu_d =\mu_s\,,\quad \mu_{\mu}=\mu_e\,.
\end{eqnarray}
The baryon (neutron) chemical potential is related to the quark
chemical potentials as
\begin{eqnarray}
\mu \equiv \mu_n =2\mu_d +\mu_u\,.
\end{eqnarray}

In Ref.~\cite{Glen92} it was suggested that in
multi-component systems the first-order phase transition proceeds
through a mixed phase. In our case this would imply a coexistence
of quark and hadronic matter in a certain density range. In the
mixed phase both quark and hadron subsystems are charged and
electro-neutrality is realized only globally (see also Refs
\cite{G97,Schrtler00}). Coulomb and surface effects were
disregarded in those works. The latter effects, however, play a
crucial role.  As argued in Ref.~\cite{VYT}, the region of the
hadron--quark mixed phase is very narrow, if not absent at all,
due to rather high surface-tension and charge-screening effects on
the quark--hadron boundary.
 Therefore, we can
use  Maxwell construction in order to describe the quark--hadron
phase transition\footnote{Even when the mixed phase does exist,
as in the case of the kaon-condensate phase transition, the
Maxwell construction proves to be effectively relevant due to
essential role of the charge-screening effects, cf.
\cite{MTVTC}.}.
This implies the local electro-neutrality
\begin{eqnarray}
\frac{2}{3}n_u -\frac{1}{3}n_d -\frac{1}{3}n_s -n_e -n_{\mu}=0,
\end{eqnarray}
where $n_u$, $n_d$, $n_s$ are quark densities, $n_e$ is the electron density
and  $n_{\mu}$ is the muon density.
The lepton contribution to the pressure is given by
\begin{eqnarray}
P_l =\sum_{i=e,\mu}\left[ n_{i}\mu_{i}-\int^{p_{Fi}}_{0} \frac{p^2
    dp\sqrt{p^2 +m_{i}^2}}{\pi^2}\right]\,,
\end{eqnarray}
where $m_e$ and $m_{\mu}$ are lepton masses,
$\mu_{i}=\sqrt{p^2_{Fi} +m_{i}^2}$, $e$ and $\mu$ densities are
\begin{eqnarray}
n_i =\frac{(\mu_{i}^2 -m_{i}^2
  )^{3/2}}{3\pi^2}\theta (\mu_{i} -m_{i}),
\end{eqnarray}
and $\theta (x)$ is the ordinary step function.

%
%------------------------------------------------------------------------
\begin{figure*}
\includegraphics[height=60mm,clip]{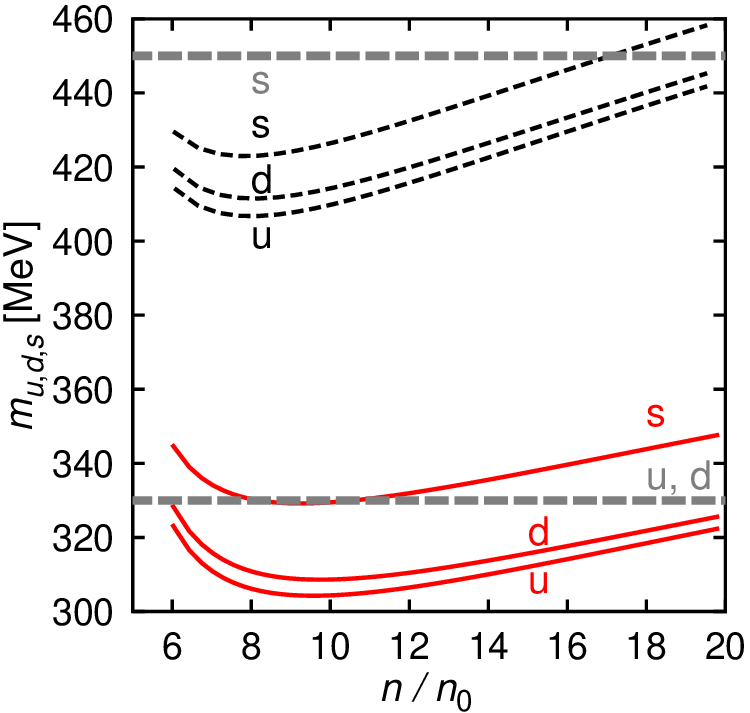}\quad
\includegraphics[height=60mm,clip]{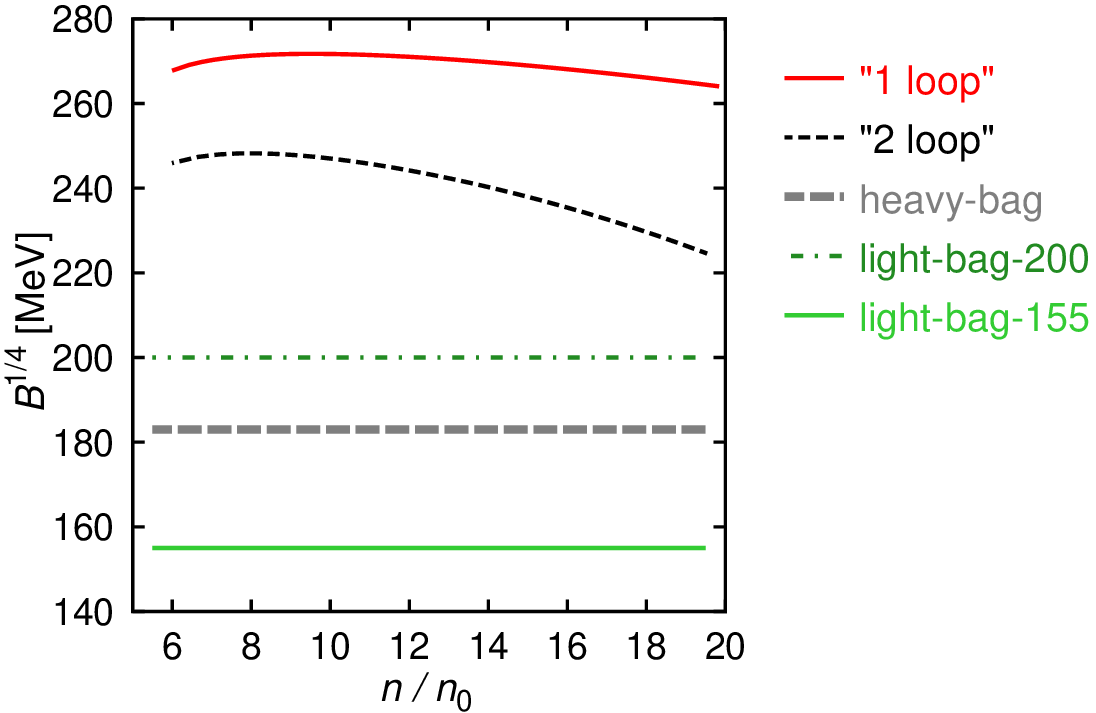}
\caption{(Color online)
Effective quark masses (left panel) and bag
  parameters (right panel) for the DQ models (curves),
  cf. Eq. (\ref{U_g}), and bag models (horizontal lines)
 as  functions of $n/n_0$ for $T=0$ at $\beta$ equilibrium.
Notation is the same as in Fig.~\ref{p0_fig}.
} \label{mass-q}
\end{figure*}
%------------------------------------------------------------------------

Now let us specify the hadronic EoS.
The Urbana-Argonne EoS~\cite{APR98} $A18+\delta v+UIX^*$ complies with the rich experimental
information available for $n\lsim n_0$ ($n_0$ is the nuclear saturation
density). For that reason this EoS is considered to be the
most realistic EoS at present. However, it uses a non-relativistic
potential and hence violates causality at $n \gsim 4\,n_0$.
Ref.~\cite{HHJ} suggested a simple analytical parameterization
(below referred as the HHJ EoS) of $A18+\delta v+UIX^*$ EoS valid
for $n\lsim 4 n_0$, which respects causality
at higher densities.
%%We denote this parameterization as the HHJ EoS.
On the other hand, as an extrapolation to higher densities,
relativistic mean field (RMF) models are practical. However all
RMF models have an unpleasant feature. They produce  a
large fraction of protons in neutron star matter that  permits
a very efficient cooling process, i.e. the direct Urca (DU) process
$n\rightarrow pe\bar{\nu}$, for rather low densities $<3n_0$.
First, this disagrees with predictions of the mentioned
$A18+\delta v+UIX^*$ EoS, where the DU process is allowed only for
$n>5n_0$. Second,  a low threshold density $n_{\rm crit}^{\rm
DU}\lsim 3~n_0$ seems to be at odds with a neutron star cooling scenario
\cite{BGV04} predicted by the Urbana-Argonne EoS.

Another problem is that rather massive neutron stars may exist. Recent
measurements of neutron star masses in binary compact systems yielded
$M_{\rm J0751+1807}=2.2\pm 0.2 M_{\odot}$ \cite{S04} at the $95\%$ confidence
level.  Even assuming the highest value of the  uncertainty of this
measurement one can conclude that $M_{\rm J0751+1807}>1.6 M_{\odot}$ (within $2\sigma$).
Some models with a so-called soft hadron EoS do not satisfy this requirement. Note that
the Urbana--Argonne EoS yields  the limiting mass of a neutron star $\simeq
2.2 M_{\odot}$. With the relativistic HHJ EoS the limiting neutron star mass
decreases slightly below $\simeq 2~M_{\odot}$. Possible phase transitions to
the either pion or kaon condensation, hyperonization, quark matter,
etc. soften the EoS, that may significantly  decrease the limiting mass of the star, cf. Ref. \cite{glen}.

Ref. \cite{KV04} suggested a solution of both problems within an effective RMF
model with field dependent effective hadron masses and coupling constants.
We will use two models from this work.
One model, ``hadron-MW2'' (MW(nu)($z=2.6$) in the notation
of Ref. \cite{KV04}), fits the HHJ EoS for all densities yielding the threshold
density for the DU
process $\simeq 5.2n_0$.
Another model, ``hadron-MW1''
(MW(nu)($z=0.65$)  in the notation
of Ref. \cite{KV04}) fits the HHJ EoS for $n\lsim 2n_0$ but  allows to
increase  the 
limiting mass of a neutron star due to a stiffening of  the EoS at a higher
density. The threshold of the DU process is  $\simeq 4n_0$ which
does not contradict to the above mentioned description of the star cooling.
 Thus, the ``hadron-MW1'' model possesses the same advantages, as the
Urbana--Argonne EoS and
the ``hadron-MW2'' one, but in addition allows to increase the limiting mass
of the star. The pressure of the ``hadron-MW1'' model
as a function of the baryon chemical potential is lower as
compared to that of the   ``hadron-MW2'' model, which is in favor of the phase
transition to the quark state at a smaller density. In order to
keep the considerations simple,
we use models that disregard the possibility of $\rho^-$ condensation
%at over-critical densities
discussed in Ref. \cite{KV04}.

In Fig.~\ref{mass-q} (left panel) we show effective masses of
quarks  at $T=0$ as a function
of the baryon density in units of the normal nuclear matter density $n_0
\simeq 0.16$ fm$^{-3}$.
Horizontal lines represent quark masses of
the ``heavy-bag'' model, see Table \ref{tab:bagpar}.
Self-consistent calculations of the DQ model
produce a much smaller difference between strange and non-strange quark masses
compared to $120\div 150$~MeV, which is usually accepted by bag
models. In the DQ model, quark
masses slightly decrease with increasing density up to $(8\div 9)~n_0$.
For higher  densities the behavior reverses and the masses undergo a
slight increase.
Note that such a density behavior of the quark masses is quite different from
that obtained in the NJL model
\cite{Klevansky}. In the
latter case for zero temperature  the quark
masses drop with increasing density,
reaching  zero at a moderate value
of the baryon density $n_c$,  they also drop to zero
at $T=T_C$.  Therefore it seems doubtful that the
NJL model  could reproduce the lattice data at least
near and above  $T_C$; similarly we have shown that the ``light-bag'' models
fail.

%------------------------------------------------------------------------
\begin{figure*}
\includegraphics[height=59mm,clip]{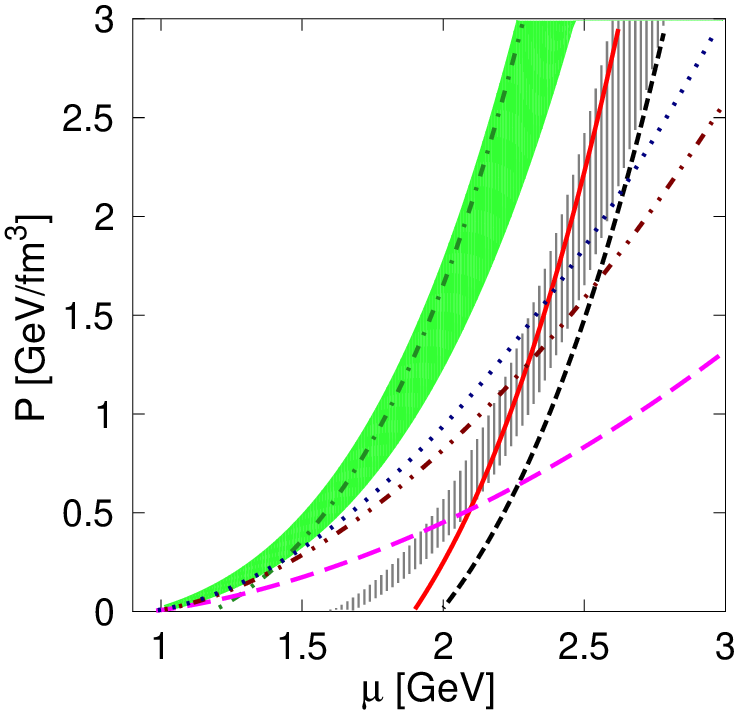}\quad
\includegraphics[height=59mm,clip]{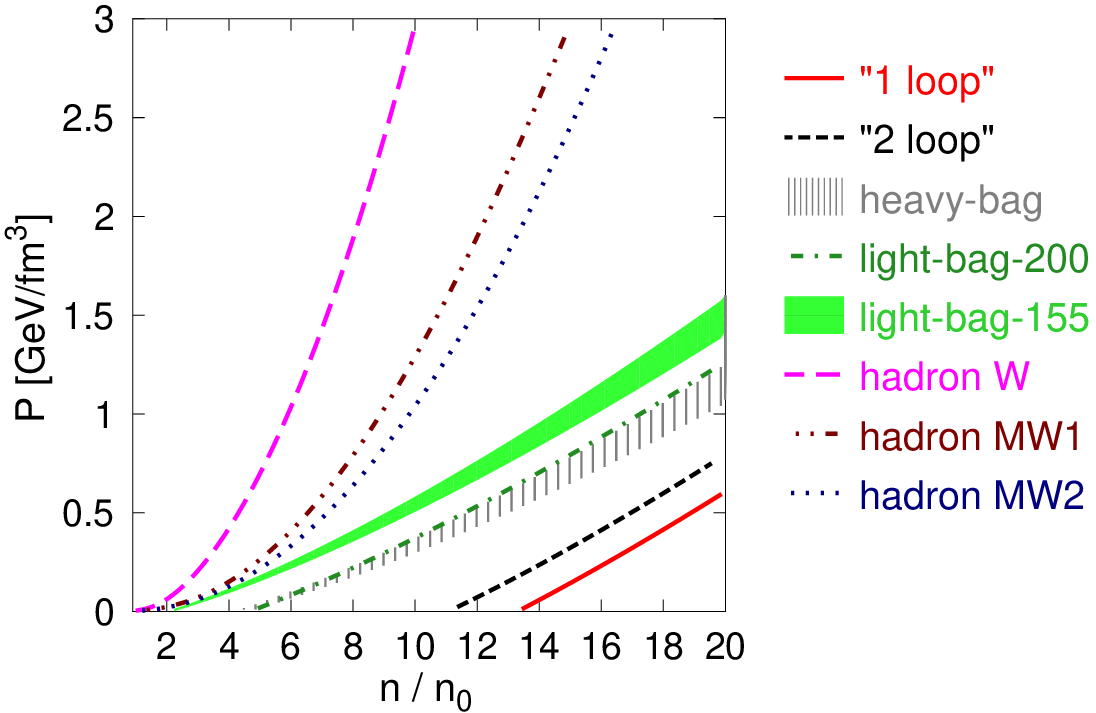}
\caption{(Color online)
Pressure as  a function of the chemical potential (left
  panel) and $n/n_0$ (right panel) for $T=0$ at $\beta$ equilibrium.
Notation for the quark models is the same as in Fig.~\ref{p0_fig}.
Band boundaries for the ``heavy-bag'' and ``light-bag-155'' model
correspond to $c=0$ and $c=0.3$: from  top to bottom in left
panel and from bottom  to top in right panel. Hadronic EoS of the
conventional Walecka model (``hadron W'') is displayed by long-dashed line. Two
versions of the hadronic model of Ref. \cite{KV04}, i.e. ``hadron-MW1''
and  ``hadron-MW2'', are presented by dashed-dotted-dotted and dotted
lines, respectively.
} \label{P-star}
\end{figure*}
%------------------------------------------------------------------------

Fig.~\ref{mass-q} (right panel) demonstrates the behavior of the
effective bag parameter $B^{1/4}$ as a function of density at zero
temperature. The $B$ values prove to be very high in the DQ model,
whereas the bag parameter of the ``heavy-bag'' model is in the
range of ordinarily used values. Implications of a
temperature-density dependent bag parameter, in particular for
quark stars, have been earlier considered in Ref.
\cite{Blaschke98,Schrtler00}. 

\subsection{Do hybrid stars exist?}

In Fig.~\ref{P-star} we present the pressure as a function of the
baryon chemical potential (left panel)  and as a function of the baryon
density (right panel)  for the models considered above.
Realistic hadronic RMF EoS's are represented by the ``hadron-MW2'' and
``hadron-MW1'' models discussed above.
The latter EoS is stiffer, as is seen from the right panel.
As a reference model, the stiffest RMF EoS of the original Walecka model
(``hadron-W'') is also demonstrated. The crossing
points of the hadronic and quark curves on the $(P,\mu)$ plane
(left panel) show critical values ($\mu_c ,P_c$) for
occurrence of the first-order phase transition from the hadronic
to the quark phase. The latter transition arises,  if for $\mu
>\mu_c$ the pressure of the quark phase is higher than that for
the hadron phase. The corresponding  critical densities $n_c^h$
and $n_c^q$ at the Maxwell construction, $n_c^h < n_c^q$, are
listed in Table~\ref{tab:ncrit}.

As seen, all lattice-QCD-motivated models (DQ and ``heavy-bag'') of
the quark phase matched with realistic RMF models
(either  ``hadron-MW1'' or ``hadron-MW2'') of the  hadronic phase
predict the onset of the phase transition at very high critical densities
(typically $n_c^h\sim 10\,n_0$).
Note that both
``hadron-MW1'' and ``hadron-MW2'' models do not permit neutron
stars with central baryon densities $\geq 8~n_0$, see Ref. \cite{KV04}.
{\em{Thus, our lattice-QCD-motivated models (DQ and ``heavy-bag'')
matched with realistic RMF models
(either  ``hadron-MW1'' or ``hadron-MW2'')
do not allow the existence of hybrid stars.}}
The stiffer the hadron EoS, the lower the critical density
$n_c^h$ is. However, the central density of the limiting mass star also
decreases. In the case of the DQ and ``heavy-bag'' models matched with
the ``hadron-W'' model representing unrealistically stiff
EoS (i.e the stiffest EoS among those still used in
  calculations),
the typical critical density $n_c^h \approx 4n_0$.  The critical
masses $M(n_{cent}=n_c^h$) for the occurrence of the hybrid stars
are listed in Table~\ref{tab:nsmass}. Three hadron-quark model
combinations, i.e. ``hadron-W''+``1-loop'' and
``hadron-W''+``heavy-bag'' with $c=0$ and $c=0.3$,
predict the existence of hybrid stars with
$M>2.7~M_{\odot}$. These mass values are
only slightly below the limiting mass.
%{\em{
The hadron-quark
combination ``hadron-W''+``2-loop'' predicts no hybrid stars.
%}}

\begin{table*}
\begin{ruledtabular}
\begin{tabular}{ccccccc}
%hadr.$\backslash$ quark  & ``1-loop'' & ``2-loop'' &  h-bag($c=0$) &  h-bag(0.3)& l-bag-155(0.3)   & l-bag-200(0)
hadr.$\backslash$ quark  & ``1-loop'' & ``2-loop'' &  heavy-bag($c=0$) &  heavy-bag(0.3)& light-bag-155(0.3)   & light-bag-200(0)
\\ \hline
W   &  4.4/20.3 &  4.8/19.3 &  3.9/12.1 &  4.4/12.6  & 1.3/2.6 & 2.2/6.4 \\
MW1 &  9.8/27.7 & 11.2/29.7 &  9.3/22.0 &  10.8/24.1 & 3.7/4.3 & 4.4/7.1 \\
MW2 & 12.2/31.3 & 14.1/34.5 & 11.5/25.1 &  13.7/29.1 & 5.8/7.1 & 5.4/9.1 \\
\end{tabular}
\end{ruledtabular}
\caption{ The
hadronic $n_c^h$ and quark $n_c^q$ critical densities (in units $n_0$)
according to the Maxwell construction.
%Labels of the EoS's are slightly shorten.
}
\label{tab:ncrit}
\end{table*}

The ``light-bag-155($c=0.3$)'' and ``light-bag-200($c=0$)'' models result in
critical densities in a range which is usually considered in papers
devoted to hybrid stars.
For the ``light-bag-155($c=0$)'' model matched with any of  the
considered hadron models, quark matter is always more stable
than hadronic matter (the quark pressure is higher), and hence quark stars of
an arbitrary size are possible.
In this case only a softer (compared to those we have used) hadronic EoS
could allow the possibility of hybrid star existence.

\begin{table*}
\begin{ruledtabular}
\begin{tabular}{ccccccc}
%hadron$\backslash$ quark  & ``1-loop'' & ``2-loop''&  h-bag($c=0$) &  h-bag(0.3)& l-bag-155(0.3)   & l-bag-200(0)
hadr.$\backslash$ quark  & ``1-loop'' & ``2-loop'' &  heavy-bag($c=0$) &  heavy-bag(0.3)& light-bag-155(0.3)   & light-bag-200(0)
\\ \hline
W   &  2.7& -- &  2.8& 2.7& 0.6& 2.1 \\
MW1 &  -- & -- &  -- & -- & 1.6& 1.8 \\
MW2 &  -- & -- &  -- & -- & 1.8& 1.7 \\
\end{tabular}
\end{ruledtabular}
\caption{ Masses  of neutron stars $M$ (in units $M_{\odot}$)
corresponding to the critical densities
in the hadron phase, $n_c^h$, listed in Table~\ref{tab:ncrit}.
%%and quark phases (Tab.~2).
%Labels of the EoS's are slightly shorten.
}
\label{tab:nsmass}
\end{table*}

\subsection{Quark stars}

Now we turn to the question of the possible existence of pure
quark stars.  As argued in Ref.~\cite{AFO}, the quark star may
be surrounded by only a tiny crust separated from the core by a
strong electric field. Therefore, we will neglect  contribution of
the crust. Then the star mass and its radius result from the
Tolman-Oppenheimer-Volkoff equation (in the standard notation,
cf. \cite{MS})
\begin{eqnarray}
\frac{d P}{d r}&=&-\frac{G\,\varepsilon\,m}{r^2}\,
\frac{1+P/\varepsilon}{1-2G\ m/r}
\,\left(1+\frac{4\pi\,P\,r^3}{m}\right),
\\ \nonumber
\frac{d m}{d r}&=&4\,\pi\,r^2 \varepsilon
\label{TOV}
\end{eqnarray}
which is solved for a given $\varepsilon (P)$ dependence
with initial conditions $m(0)=0$ and $P(0)=P_{\rm cent}$, where
$P_{\rm cent}$ is a
 central pressure corresponding to a central density $n_{\rm centr}$.
The latter is the input parameter of the calculation. For the sake
of simplicity, we neglect effects of possible rotation of the
star, which however may  affect the mass-radius relation of the
newly born stars and also influence the dynamics of the possible
phase transition of the neutron star to the hybrid or the quark
star, cf. \cite{CGPB}. 
%------------------------------------------------------------------------
\begin{figure}
\includegraphics[height=55mm,clip]{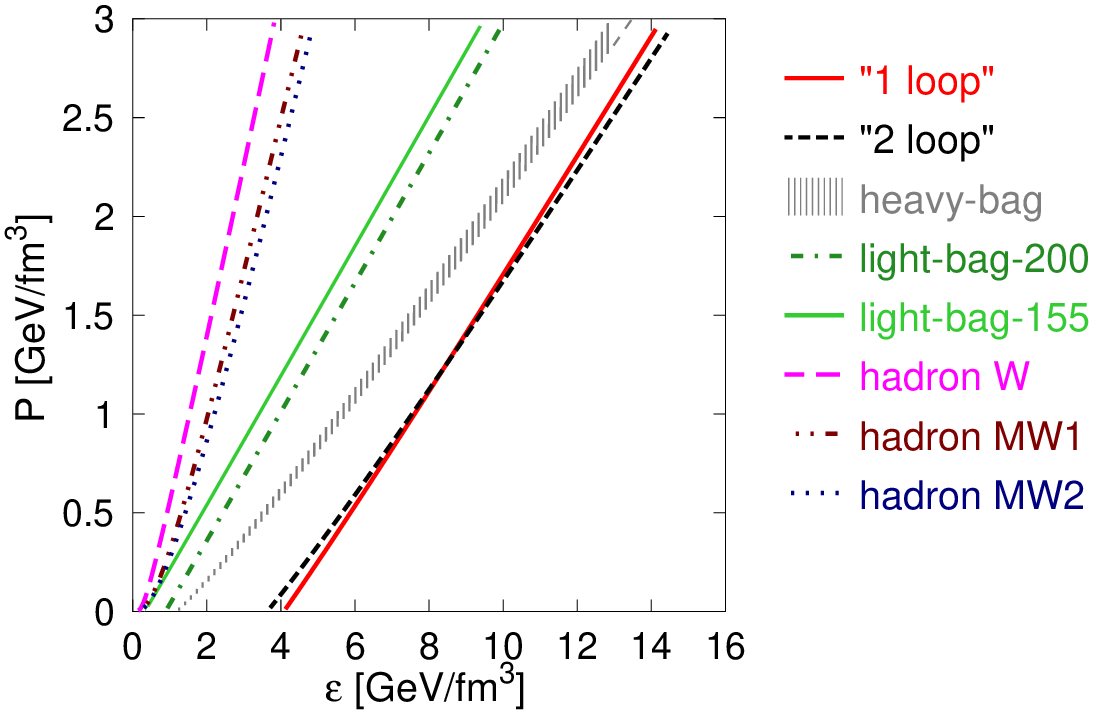}
\caption{(Color online)
Pressure versus energy density at $T=0$ in
  $\beta$-equilibrium for various EoS's. Band boundaries for the
  ``heavy-bag'' model correspond $c=0$ and $c=0.3$ (from bottom to top).
Notation is the same as in Figs~\ref{p0_fig} and \ref{P-star}.
} \label{P-E-star}
\end{figure}
%------------------------------------------------------------------------

The pressure  as a function of the energy density is shown in
Fig.~\ref{P-E-star} for the models under consideration. We see
that for all the quark models these dependencies can be well
parameterized with appropriate accuracy by straight lines
$\varepsilon =a P+b$, with $b=4\widetilde{B}$, where
$\widetilde{B}$ has the meaning of an effective bag parameter, cf.
Ref. \cite{Pesh00,Pesh00b}. For  the ``massless-bag'' model, i.e.
with all quarks being massless, we would obtain $a=3(1-c)$,
$b=4B$.
 Values of the effective bag parameter $\widetilde{B}$ and
 slop-coefficients ``$a$'' are listed in Table~\ref{tab:pepar}.
Note that the quasiparticle model of Refs \cite{Pesh00,Pesh00b}
resulted in $3.1\leq a\leq  4.5$ and $\widetilde{B}^{1/4} > 200~$MeV (typically
$\widetilde{B}^{1/4}\simeq (230\div 240)$~MeV). Within our
lattice-QCD-motivated models (DQ and ``heavy-bag'') we obtained
essentially higher values: $a =5\div 6$ for and
$\widetilde{B}^{1/4}=212\div 304$ MeV.
As seen, the  $\varepsilon (P)$
lines are quite different for the ``light-bag'',
``heavy-bag'' and  DQ models. Due to that one may expect quite
different gravitational properties of objects described by these models.

\begin{table*}
\begin{ruledtabular}
\begin{tabular}{cccccccc}
%coef.$\backslash$ model& ``1-loop'' & ``2-loop''&  h-bag($c=0$) &  h-bag(0.3)& l-bag-155(0)   & l-bag-155(0.3) & l-bag-200
Model
%coef.$\backslash$ quark
& ``1-loop'' & ``2-loop'' &  heavy-bag($c=0$) &  heavy-bag(0.3)& light-bag-155(0)   & light-bag-155(0.3)   & light-bag-200(0)
\\ \hline
$a$  &  5.01 &  5.97 &  5.89& 5.41& 3.11& 3.09& 3.15 \\
$\widetilde{B}^{1/4}$, MeV &  304& 293 &  216& 212& 159& 159& 202\\
\end{tabular}
\end{ruledtabular}
\caption{
Coefficients of the linear interpolation of the $\varepsilon (P)$ dependence.
% $\widetilde{B}^{1/4}$ is measured in MeV.
}
%Labels of the EoS's are slightly shorten.
\label{tab:pepar}
\end{table*}

%------------------------------------------------------------------------
\begin{figure*}
\includegraphics[height=64mm,clip]{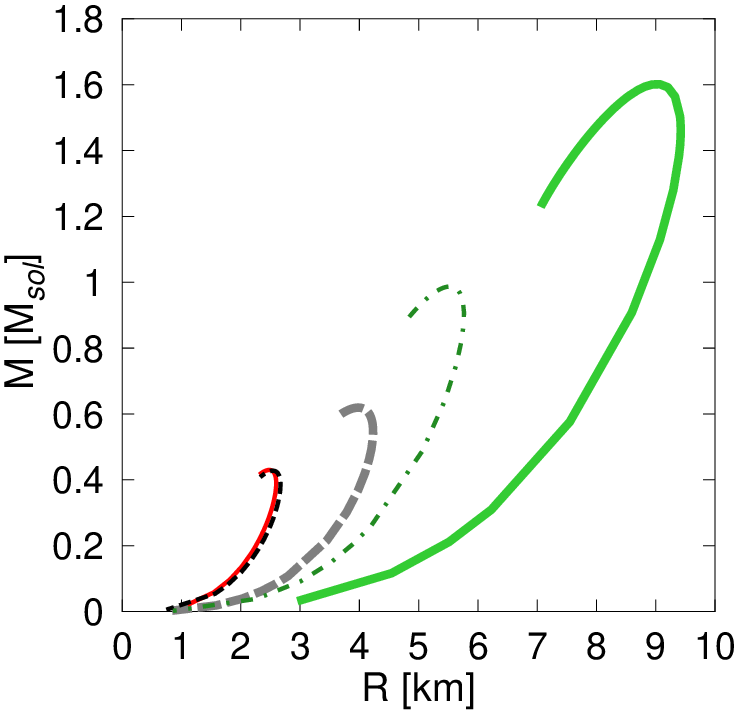}\quad
\includegraphics[height=64mm,clip]{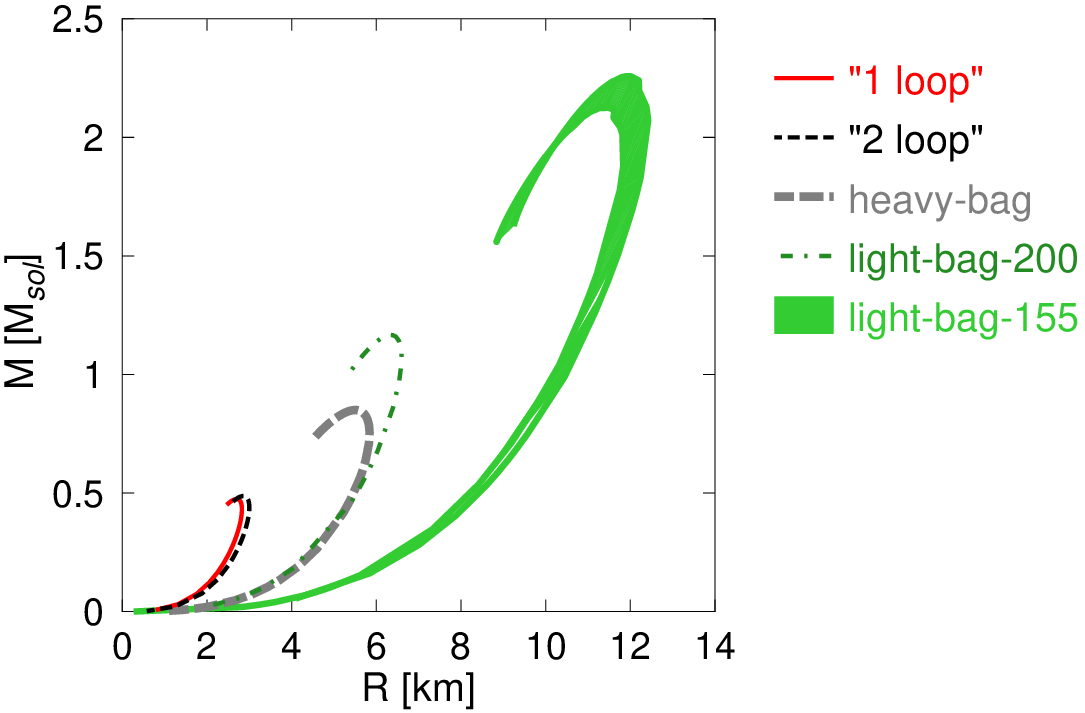}
\caption{(Color online)
%%Notation is the same as in Fig. \ref{p0_fig}.
Quark star mass -- radius relation of pure quark stars for the normal quark
phase with $\Delta =0$ (left panel) and the CFL  quark
phase with $\Delta =100~$MeV (right  panel). Notation is the same as
in previous figures. Band boundaries $c=0$ and $c=0.3$, from bottom to
top.
} \label{NS}
\end{figure*}
%------------------------------------------------------------------------

%------------------------------------------------------------------------
\begin{figure*}
\includegraphics[height=64mm,clip]{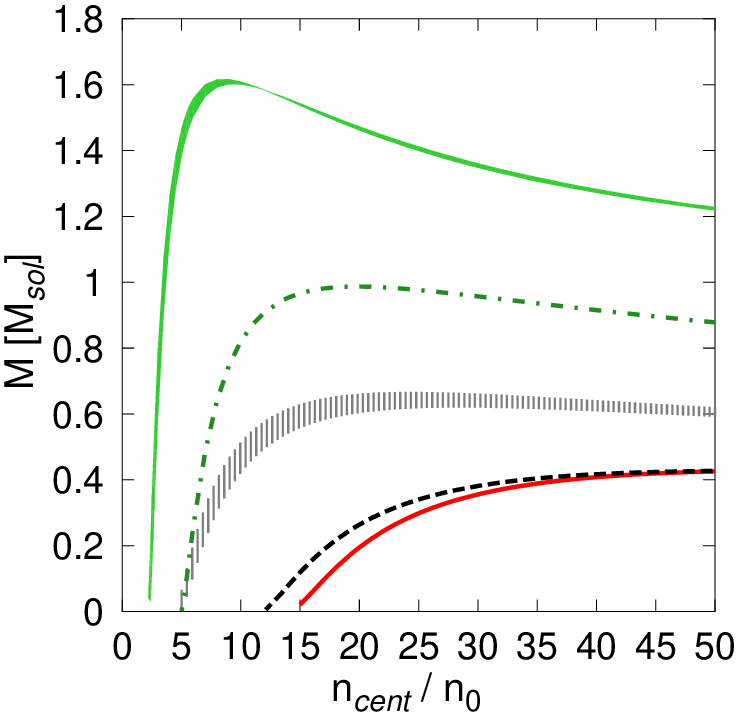}\quad
\includegraphics[height=64mm,clip]{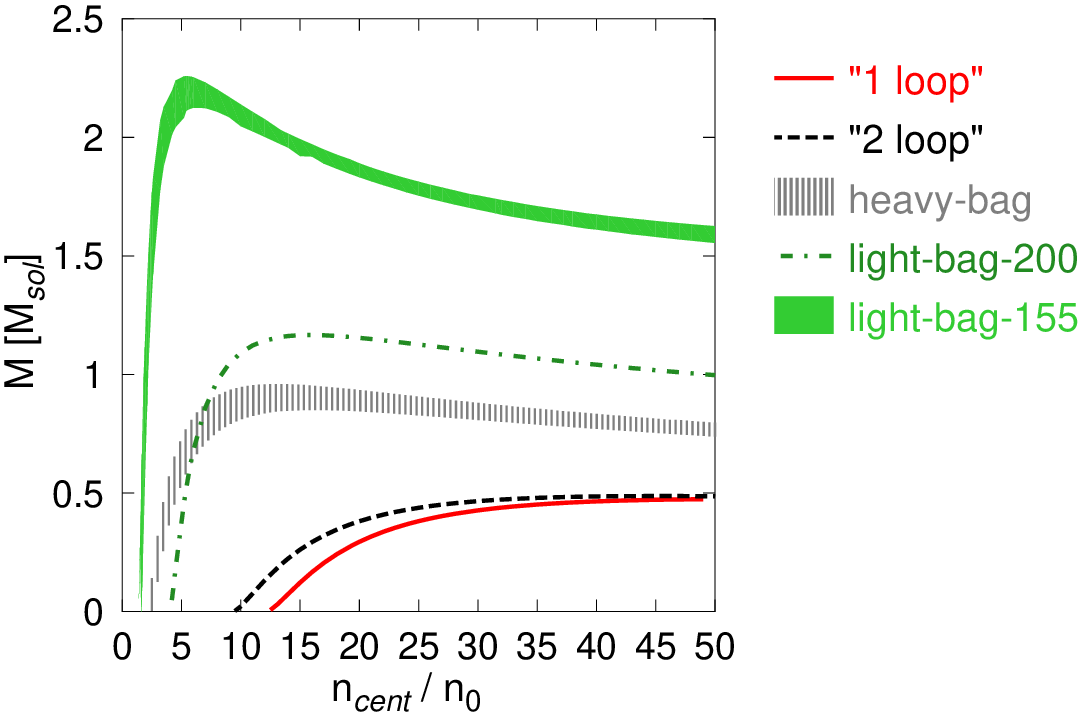}
\caption{(Color online)
Quark star mass --  central density relation. Notation is the same as
in Fig.~\ref{NS}.
%Band boundaries $c=0$ and $c=0.3$: from bottom to top.
} \label{NS1}
\end{figure*}
%------------------------------------------------------------------------

In Fig.~\ref{NS}  we show mass-radius ($M$--$R$) relation for our
models of quark stars,  $M\equiv m(r=R)$. Two extreme cases
are demonstrated: quark matter in the normal phase (left panel)
and the CFL phase of quark matter with a high pairing gap $\Delta
=100$~MeV (right panel). In the latter case the contribution of
pairing to the energy density is the largest. For other
superconducting phases, the results will range in between these
two extremes presented here. Note that occurence of high $n_c^q$
values in our DQ and ``heavy-bag'' models, see Table
\ref{tab:ncrit}, implies that the CFL phase takes place in these
cases.
Fig.~\ref{NS1} demonstrates the quark star mass as a function of the
central baryon density, again for the normal phase with $\Delta =0$
(left panel) and the CFL phase with $\Delta =100$~MeV (right panel).
The limiting masses and central densities are listed in
Table~\ref{tab:limmass} for normal
quark matter and in Table~\ref{tab:limmasspair} for the CFL quark
matter. Note that CFL strangelets were discussed in Ref. \cite{madsen}.

\begin{table*}
\begin{ruledtabular}
\begin{tabular}{cccccccc}
%norm. $\Delta =0$   & ``1-loop'' & ``2-loop''&  h-bag($c=0$) &  h-bag(0.3)& l-bag-155(0)   & l-bag-155(0.3) & l-bag-200
%$\Delta =0$
Model
%{\bf normal phase} $\backslash$ quark
& ``1-loop'' & ``2-loop'' &  heavy-bag($c=0$) &  heavy-bag(0.3)& light-bag-155(0)   & light-bag-155(0.3)   & light-bag-200(0)
\\ \hline
$M_{lim}$   &  0.43&  0.43 &  0.62&  0.67& 1.60& 1.62& 0.99 \\
$n_{centr}$ &  61.0&  55.6 &  28.3&  24.3& 9.42& 8.54&  19.8 \\
\end{tabular}
\end{ruledtabular}
\caption{
Limiting quark star masses (in units $M_{\odot}$) and corresponding central
densities (in units $n_0$) for the normal phase ($\Delta =0$) of
quark matter.
%Labels of the EoS's are slightly shorten.
}
\label{tab:limmass}
\end{table*}

\begin{table*}
\begin{ruledtabular}
\begin{tabular}{cccccccc}
\hline
%CFL,$\Delta =100$~MeV   & ``1-loop'' & ``2-loop''&  h-bag($c=0$) &  h-bag(0.3)& l-bag-155(0)   & l-bag-155(0.3) & l-bag-200
Model
%{\bf CFL phase} ,$\Delta =100$~MeV $\backslash$ quark
& ``1-loop'' & ``2-loop'' &  heavy-bag($c=0$) &  heavy-bag(0.3)& light-bag-155(0)   & light-bag-155(0.3)   & light-bag-200(0)
\\ \hline
$M_{lim}$  &   0.47&  0.49 &  0.85 &  0.96& 2.13& 2.26& 1.17\\
$n_{centr}$ &  53.2&  46.9 &  16.1 &  13.0& 6.41& 5.02& 15.8\\
\end{tabular}
\end{ruledtabular}
\caption{The same as Table~\ref{tab:limmass} but for the CFL phase of
  quark matter with $\Delta =100$~MeV.}
\label{tab:limmasspair}
\end{table*}

In the case
of ``light-bag-155'' model  the  pairing may
substantially affect the EoS and hence quark star masses. The
difference between
normal and CFL quark star masses  increases up to
$0.6~M_{\odot}$ in the region $M\sim M_{lim}$.
The limiting mass is $M_{lim}\sim (1.6\div 2.3) M_{\odot}$
and the radius is $\sim 9\div 12$~km.
Thus, for these models the $M$--$R$ and $M$--$n_{\rm cent}$
relations are similar to those one usually uses for pure hadronic
EoS's, in agreement with the statement of Ref. \cite{ABPR}.
For the ``light-bag-200($c=0$)'' model
the limiting mass is $M_{lim}\sim (0.9\div 1.2)~M_{\odot}$ and the
radius is $\sim 6$~km. This limiting mass
is smaller than several well measured masses of pulsars. Thus we
may conclude that  the ``light-bag-200''
model does not describe ordinary pulsars but rather another family of
compact objects. The effect of the
pairing is less pronounced here  than for the ``light-bag-155''  model.
Let us remind that the ``light-bag-155($c=0.3$)'' and
``light-bag-200($c=0$)'' models
matched with either ``hadron MW1'' or ``hadron MW2'' hadronic models
also predict existence of
hybrid stars. Had we assumed a softer hadron EoS, the corresponding
hybrid stars would
become impossible and we would only have pure quark stars.

For the ``heavy-bag'' model and especially for the DQ model, the masses
and radii are still smaller and the interior region, denser.
For the ``heavy-bag'' model
the limiting mass is $M\sim (0.6\div 1)~M_{\odot}$ and the radius is $(\sim
4\div 6)$~km. These configurations are similar to those previously studied in
the model of Ref. \cite{Pesh00b}. For the DQ model the limiting mass is $M\sim
(0.4\div 0.5)~M_{\odot}$ and the radius is $(\sim 2.5\div 3)$~km.
The central density is $\sim (50\div 60) n_0$.
Such dense quark matter is probably
realized in the CFL phase that has peculiar cooling properties.
Small and dense objects might be of interest with respect to MACHO events
\cite{frada}.

\section{Conclusions}

Models for deconfined quark matter  are extensively used in astrophysical
applications. The discussion of hypothetical hybrid and quark stars is a
rather hot topic.
In this paper we emphasize that it is essential for a deconfined
quark model to be constrained by
the lattice QCD results, which became recently available at finite
baryon chemical potential.
Rapidly progressing lattice calculations may essentially restrict models
used for the description of cold and dense quark matter. In this paper
we are trying to make a step in this direction. At this time, we would
like to draw attention  to problems with traditionally used simple
standard bag and, probably, NJL  models. They are  not able to reproduce
the lattice data without significant  modifications.

We have considered the dynamical quasiparticle  model~\cite{I04}
and the heavy-bag model, which fit the lattice results.
The heavy-bag model uses heavy masses of quarks and gluons to
reproduce the lattice results well.
We have demonstrated, that these models, being matched with realistic
hadron models of EoS, rule out the existence of hybrid stars. This
implies that families of ordinary neutron stars and
pure quark stars are well separated. Within
these models the quark stars have low masses ($M\lsim 1~M_{\odot}$),
are dense ($n\gsim 10n_0$) and very compact ($R\lsim 6$~km).
This result complies with that obtained in Ref. \cite{Pesh00b} by means
of a different extrapolation of lattice data to cold dense baryonic
matter. If such a quark
star existed in a binary system (together with a white dwarf,
a neutron star, or together with another quark star), the
orbiting would be different from that experimentally known
for both white dwarf--neutron star and double neutron star
systems. Moreover, such dense quark matter may be in the
color-flavor-locked superconducting phase and hence possess
peculiar cooling
properties. Other properties of these pure quark stars also
differ from those of neutron stars that may give a
chance to distinguish them experimentally, e.g., see Ref. \cite{madsen}.
%(certainly if above discussed models were indeed relevant).

Another interesting question arises, what could be a possible
formation mechanism of such compact and dense quark stars? For the
birth of a quark star, a hybrid star configuration is first required
 in order to produce a quark core either during supernova
collapse or during the long-term accretion of the matter in a
binary system, or in the course of the merging of two neutron
stars. The nucleation dynamics of quark matter in a cold neutron
star was considered recently in Ref.~\cite{Bomb04}. Then also a
mechanism to blow off the hadron shell is needed, e.g., like that
discussed in Ref.~\cite{1orderoff} for a strong first-order
phase transition.

In the present paper the models for deconfined matter, which are
fitted to lattice predictions, allow neither hybrid stars nor
quark self-bound objects (strangelets). We have demonstrated that
only the unrealistically stiff original Walecka  model
(``hadron-W'') matched with either the DQ or ``heavy-bag'' models
may allow for a small quark core. Thus, we actually see no
appropriate mechanism to form quark stars that are described by
our DQ and heavy-bag models. Temperature, rotation, and neutrino
trapping effects should be included. They may stiffen the hadron
EoS~\cite{Prakash97} and perhaps may help to resolve the problem.
Some exotic ways to produce quark stars through either evolution
of  white dwarfs or inflation stage at the quark--hadron phase
transition are discussed in Refs \cite{frada,Borghini}. Certainly,
this problem still needs further consideration.

It is necessarily to mention uncertainties in the present considerations
and the conclusions derived. First of all, they are connected with
lattice data \cite{latEoS}, which we extrapolate to the domain of cold
and dense
baryonic matter. These data  obtained on a discrete lattice are only
poorly extrapolated to the continuum limit. Another problem with the
lattice data consists in the poor reproduction of the chiral limit. As we
have mentioned already, the pion turns out to be very heavy in lattice
calculations, e.g. $m_{\pi}\simeq 770$~MeV \cite{Redlich}. All
this may result in
certain errors in the parameters of our effective models, deduced from
these data. We have tried to estimate this kind of uncertainties by
applying different extrapolations from the lattice data,
i.e. the ``1-loop'' and ``2-loop'' versions of the DQ model and the
``heavy-bag'' model. Indeed, whereas the ``2-loop'' version closely
reproduces the lattice data \cite{latEoS}, the ``1-loop'' version
slightly (by 10\%) misfits its normalization, thus leaving room for
uncertainty in the extrapolation to the continuum and chiral
limits. As for the
``heavy-bag'' model, it covers the lattice data by a rather broad band
(due to variation of $c$).

On the other hand, the grounds for the standard bag model, in our case
represented by the ``light-bag-155'' and ``light-bag-200'' models, are not
entirely reliable. Indeed, the bag parameters of standard bag models are
fitted to reproduce the masses of
hadrons, i.e. few-body systems, and then applied to describe
matter. It could happen that the situation here is similar to that of
Walecka model for nuclear matter. If we fit the Walecka model to
reproduce the deutron or helium and then apply it to $^{208}$Pb, we
surely get nonsense. Therefore, predictions of these models for the
EoS should be treated with great care.

Theoretical uncertainties in our predictions should be mentioned
as well. The region of  the vicinity of $T_C$ deserves
special discussion. Besides the above mentioned uncertainties of
the lattice calculations, as it is known, in the (2+1)-flavor case
the phase transition at small chemical potentials is of the
cross-over type \cite{latEoS0}. 
This implies that the hadronic and quark--gluon
degrees of freedom essentially interact in this region, e.g., see
Ref. \cite{SZ03}. This is why we cannot pretend to reproduce the
lattice data in this region by means of only quark--gluon degrees
of freedom incorporated in the DQ model. Our aim was to fit these
data beyond this vicinity region and then to extrapolate them to
the domain of the cold and dense baryon matter.
As it is commonly
expected, the phase transition in this domain is of the first
order, which implies a less severe role of the
hadron--quark--gluon interaction. Therefore, we hope that the
method of matching hadronic and quark--gluon EoS's, applied in
this domain, is reasonable.
However,
%%this kind of
the uncertainty of such an extrapolation should be
%%also
kept in mind.  We cannot also
%%, of course,
exclude that  our  models are oversimplified and some important
physics is left aside, e.g., the model suffers of a lack of the
chiral symmetry restoration,  thermal fluctuations are not
included, which may play an important role in the vicinity of the
critical temperature but die out at smaller and higher
temperatures, etc. This may somehow modify the fitting parameters
of models. Quantum fluctuations like instantons and gluon
condensate may partially survive and additionally contribute in
the low temperature region, cf. Ref. \cite{Ellis}. Thus, it might
well be that the DQ model, as well as the heavy-bag model, should
be replaced in the future by  more sophisticated and realistic
microscopically based models. Nevertheless, a certain confidence
in our results is supported by the fact that other
independent extrapolations of the lattice data to the cold dense
baryonic matter \cite{Pesh00b,Rebhan03} resulted in rather
similar conclusions. However, it would be of interest to search
for new theoretical approaches for extrapolating the EoS from
$T>190$ MeV to $T=0$ and from $\mu=530$ MeV upwards.

\acknowledgments

We are grateful to P.J. Ellis for fruitful discussions and careful reading
the manuscript.
This work was supported in part by the Deutsche
Forschungsgemeinschaft (DFG project 436 RUS 113/558/0-2), the
Russian Foundation for Basic Research (RFBR grant 03-02-04008) and
Russian Minpromnauki (grant NS-1885.2003.2).
The work of E.E.K. was supported in part by the US Department of
Energy under contract No. DE-FG02-87ER40328.


\begin{thebibliography}{99}
\bibitem{GM04}
E. Shuryak, J. Phys. G {\bf 30}, S1221 (2004);
M. Gyulassy, and L. McLerran,
e-Print Archive: nucl-th/0405013.
%
\bibitem{Itoh}
D. Ivanenko and D.F. Kurdelaidze, Lett. Nuovo Cim. {\bf{2}}, 13 (1969);
N. Itoh, Prog. Theor. Phys. {\bf 44}, 291 (1970).
\bibitem{G97}
N.K. Glendenning, {\em Compact Stars} (Springer-Verlag, 1997); F. Weber,
{\em Pulsars as Astrophysical Laboratories for Nuclear and Particle
Physics} (IOP Publishing, 1998).
\bibitem{W04}
F. Weber, astro-ph/0407155; H. Heiselberg and
V.R. Pandharipande, Ann. Rev. Nucl. Part. Sci. {\bf 50}, 481 (2000).
\bibitem{RW}
K. Rajagopal and  F. Wilczek, hep-ph/0011333.
\bibitem{bomb}
I. Bombaci, Nucl. Phys. A {\bf 681}, 205 (2001).
\bibitem{M00}
J. Madsen, Phys. Rev. Lett. {\bf 85}, 10 (2000); A. Drago, A. Lavagno, and
G. Pagliara, astro-ph/0312009.
\bibitem{bkv}
D. Blaschke, T. Kl\"{a}hn, and D.N. Voskresensky,
Ap. J. {\bf 533}, 406 (2000);
D. Page, M. Prakash, J.M. Lattimer, and A. Steiner, Phys. Rev.
Lett. {\bf{85}}, 2048 (2000);
D. Blaschke, H. Grigorian, and D.N. Voskresensky,
Astron. \& Astrophys. {\bf 368}, 561 (2001); astro-ph/0403171; astro-ph/0411619.
\bibitem{B04}
 M. Buballa, hep-ph/0402234.
%EoS calculations at finite \mu {it was earlier too, see refs. here]
\bibitem{W84}
A.R. Bodmer, Phys. Rev. D {\bf 4}, 1601 (1971);
E. Witten, Phys. Rev. D {\bf 30}, 272 (1984).
\bibitem{AFO}
C. Alcock, E. Farhi, and A. Olinto, Ap. J. {\bf 310}, 261 (1986).
\bibitem{selfbound}  J.~Schaffner-Bielich,
  %``Strangelets and strange quark matter,''
 Nucl. Phys.  {\bf A639}, 443 (1998);
C.~Greiner and J.~Schaffner-Bielich,
  %``Physics of strange matter,''
nucl-th/9801062;
J.~Madsen,
  %``Physics and astrophysics of strange quark matter,''
Lect. Notes Phys.  {\bf 516}, 162 (1999)
[astro-ph/9809032].
\bibitem{latEoS0}  F.~Karsch,
%LATTICE QCD AT HIGH TEMPERATURE AND DENSITY.
 Lect. Notes in Phys. {\bf 583}, 209 (2002).
\bibitem{latEoS} Z.~Fodor,
%Lattice QCD results at finite temperature and density
Nucl. Phys. A {\bf 715}, 319 (2003);
  F. Csikor, G.I. Egri, Z. Fodor, S.D. Katz, K.K.
Szabo, and A.I. Toth,
%EQUATION OF STATE AT FINITE TEMPERATURE AND
%CHEMICAL POTENTIAL, LATTICE QCD RESULTS.
JHEP {\bf 405}, 46 (2004).
%
\bibitem{Allton}
%The Equation of State for Two Flavor QCD at Non-zero Chemical Potential
C.R. Allton, S. Ejiri, S.J. Hands, O. Kaczmarek, F. Karsch,
E. Laermann, and C. Schmidt,
Phys. Rev. D {\bf 68}, 014507 (2003).
%%, [hep-lat/0305007].
%
\bibitem{Gavai} R.V. Gavai and S. Gupta, Phys. Rev. D {\bf 68}, 034506
  (2003).
%%, [hep-lat/0303013]
%
\bibitem{Gorenstein95} M.I. Gorenstein and S.N. Yang, Phys. Rev. D
  {\bf 52}, 5206 (1995).
%
\bibitem{Greiner} W. Greiner and D.H. Rischke, Phys. Rep. {\bf 264},
183 (1996).
%
\bibitem{Levai} P. Levai and U. Heinz, Phys. Rev. C {\bf 57}, 1879
(1998).
%
\bibitem{Pesh96}
A. Peshier, B. K\"ampfer, O.P. Pavlenko, and G. Soff, Phys. Rev. D {\bf 54},
2399 (1996);
%\bibitem{Pesh02}
%A. Peshier, B. Kampfer, G. Soff,
A. Peshier, B. K\"ampfer, and G. Soff, Phys. Rev. D {\bf 66}, 094003 (2002).
\bibitem{Pesh00}
A. Peshier, B. K\"ampfer, and G. Soff, Phys. Rev. C {\bf 61}, 045203 (2000).
\bibitem{Pesh00b}
A. Peshier, B. K\"ampfer, and G. Soff, hep-ph/0106090.
%
\bibitem{Weise01} R.A. Schneider and W. Weise, Phys. Rev. C {\bf 64},
  055201 (2001);
T. Renk, R.A. Schneider, and W. Weise, Phys. Rev. C {\bf 66}, 014902
(2002).
%
\bibitem{Rebhan03} A. Rebhan and P. Romatschke, Phys. Rev. D {\bf 68},
  025022 (2003).
%
\bibitem{Szabo03} K.K. Szab\'o and A.I. T\'oth, JHEP {\bf 306}, 008
  (2003).
%
\bibitem{Weise04} M.A. Thaler, R. Schneider, and W. Weise, Phys.Rev. C {\bf 69}, 035210
(2004).
%
\bibitem{BKS}
M. Bluhm, B. Kampfer, and G. Soff, hep-ph/041106.
%
\bibitem{I04}
Yu.B. Ivanov, V.V. Skokov, and V.D. Toneev, Phys. Rev. D {bf 71},
014005 (2005) [hep-ph/0410127].
%
\bibitem{Pisarski} E. Braaten and R.D. Pisarski, Phys. Rev. D {\bf 45}, R1827
  (1992); J. Frenkel and J.C. Taylor, Nucl. Phys. B {\bf 374}, 156
  (1992); J.P. Blaizot and E. Iancu, Nucl. Phys. B {\bf 417}, 608
  (1994);
J.O. Andersen, E. Braaten, and M. Strickland,
Phys. Rev. Lett. {\bf 83}, 2139 (1999);
Phys. Rev. D {\bf 62}, 045004 (2000);
J.O. Andersen, E. Braaten, E. Petitgirard, and M. Strickland,
Phys. Rev. D {\bf 66}, 085016 (2002);
J.P. Blaizot, E. Iancu, and A. Rebhan,
Phys. Rev. Lett. {\bf 83}, 2906 (1999);
Phys. Lett. B {\bf 470}, 181 (1999);
Phys. Rev. D {\bf 63}, 065003 (2001).
%
\bibitem{ABPR}
M. Alford, M. Braby, M. Paris, and S. Reddy, nucl-th/0411016.
\bibitem{HHJ}
H. Heiselberg, and  M. Hjorth-Jensen, ~astro-ph/9904214;
Phys. Rep. {\bf 328}, 237 (2000).
%
\bibitem{Kapusta}
J.I. Kapusta, {\em Finite-Temperature Field Theory}, Cambridge
 University Press, Cambridge, 1989.
%
\bibitem{Yndurain}
F.J. Yndurain, {\em The Theory of Quark and Gluon Interactions},
Berlin,  Springer-Verlag, 1993.
%
%\bibitem{Lebel96} M. Le Bellac, {\em Thermal Field Theory}, Cambridge
% University Press, Cambridge, 1996.
%
\bibitem{Shovk04}
I.A. Shovkovy, nucl-th/0410091.
%
\bibitem{Vfl04}
D.N. Voskresensky, Phys. Rev. C {\bf 69}, 065209 (2004).
\bibitem{Jaffe} T. DeGrand, R.L. Jaffe, K. Johnson, and J. Kiskis,
  Phys. Rev. D {\bf 12}, 2060 (1975).
%
\bibitem{Zhai}
%THE FREE ENERGY OF HOT GAUGE THEORIES WITH FERMIONS THROUGH G**5.
C.-X. Zhai and B.M. Kastening, Phys. Rev. D {\bf 52}, 7232 (1995).
%%\bibitem{Satz}
%%H. Satz, Phys. Lett., {\bf{B113}}, 245 (1982).
 %

\bibitem{Redlich}  F.~Karsch, K. Redlich, and A. Tawfik,
  Eur. Phys. J. {\bf  C29}, 549 (2003).
\bibitem{V04}{\bf}
D.N. Voskresensky,  Nucl. Phys. {\bf A744}, 378 (2004).

\bibitem{Bomb02} I.~Bombaci, eConf {\bf C010815}, 29 (2002)
[astro-ph/0201369];  astro-ph/0312452.
\bibitem{Xu02} R.~Xu,  %``Strange quark stars: A review,''
  astro-ph/0211348.
\bibitem{Schaffner04} J.~Schaffner-Bielich,
 %``Strange quark matter in stars: A general overview,''
 astro-ph/0412215.
\bibitem{Drago01} A.~Drago and A.~Lavagno, Phys. Lett. {\bf
B511}, 229 (2001).
\bibitem{Glen92} N.K. Glendenning, Phys. Rev. D {\bf 46}, (1992).
\bibitem{Schrtler00} K.~Schertler, C.~Greiner, J.~Schffner-Bielich, and
M.H.~Thoma, Nucl. Phys. {\bf A677}, 462 (2000).
\bibitem{VYT}
D.N. Voskresensky, M. Yasuhira, and  T. Tatsumi, Phys. Lett. B
{\bf 541}, 93 (2002); Nucl. Phys. {\bf A723}, 291 (2003).
\bibitem{MTVTC}
Toshiki Maruyama, T. Tatsumi, D.N. Voskresensky, T. Tanigawa, and
S. Chiba, AIP Conf. Proc. {\bf 704}, 519 (2004); nucl-th/0311076;
Toshiki Maruyama,  T. Tanigawa,  S. Chiba, T. Tatsumi, D.N.
Voskresensky, and Tomoyuki Maruyama, Prog. Theor. Phys. Suppl.
{\bf 156}, 145 (2004).
\bibitem{APR98}
A. Akmal, V.R. Pandharipande, and D.G. Ravenhall, Phys. Rev. C
{\bf 58}, 1804 (1998).
\bibitem{BGV04}
D. Blaschke, H. Grigorian, and D.N. Voskresensky, Astron. \& Astrophys. {\bf 424},
979 (2004).
\bibitem{S04}
I.H. Stairs, Science, {\bf 304},  547 (2004); D.J. Nice, E.M. Splaver, and
I.H. Stairs, astro-ph/0411207.
\bibitem{glen}
A. B. Migdal, E. E. Saperstein, M. A. Troitsky, and D.N.
Voskresensky, Phys. Rep.  {\bf 192},  179 (1990);
N.~K.~Glendenning, Phys. Rep. {\bf 342},   393 (2001).
\bibitem{KV04}
E.E. Kolomeitsev and D.N. Voskresensky, nucl-th/0410063.
\bibitem{Klevansky}
S.P. Klevansky, Rev. Mod. Phys. {\bf 64}, 649 (1992).
\bibitem{Blaschke98}
D. Blaschke, C.D. Roberts, and S. Schmidt,  Pys. Lett.  {\bf
B425}, 232 (1998); D. Blaschke, H. Grigorian, G.S. Pogosyan, C.D.
Roberts, and S. Schmidt,  Pys. Lett.  {\bf B450}, 207 (1999).
\bibitem{MS}
J.~Macher and J.~Schaffner-Bielich,
 %``Phase Transitions In Compact Stars,''
 Eur.\ J.\ Phys.\  {\bf 26}, 341 (2005).
\bibitem{CGPB}
E. Chubarian, H. Grigorian, G.S. Pogosyan, and D. Blaschke,
Astron.\& Astroph. {\bf 357}, 968 (2000); N. Glendenning and F.
Weber, Lect. Notes. Phys. {\bf 578}, 305 (2001).
\bibitem{madsen}
J. Madsen, Phys. Rev. Lett. {\bf 87}, 172003 (2001).
\bibitem{frada}
E. Fraga, R.D. Pisarski, J. Schaffner-Bielich, Phys. Rev. D {\bf 63}, 121702 (2001).
\bibitem{Bomb04} I.~Bombaci, I.~Parenti, and I. Vida\~na,
astro-ph/0402404.
\bibitem{1orderoff}
A.B.~Migdal, A.I.~Chernoutsan, and I.N.~Mishustin, Phys. Lett.  {\bf
B83}, 158 (1979); B.~K\"ampfer, Pys. Lett.  {\bf B101}, 366 (1981);
B.~K\"ampfer, Pys. Lett.  {\bf B153}, 121 (1985).
\bibitem{Prakash97}  M.~Prakash, I.~Bombaci, M.~Prakash, P.J.~Ellis,
J.M.~Lattimer, and R.~Knorren,
  %``Composition and Structure of Protoneutron Stars,''
  Phys. Rept.  {\bf 280}, 1 (1997)
\bibitem{Borghini}
N. Borghini, W.N. Cottingham, and R. Vinh Man, J. Phys.  {\bf
G26}, 771 (2000).
\bibitem{SZ03} E.~Shuryak and I.~Zahed,
%Rethinking the Properties of the Quark-Gluon Plasma at $T\sim T_c$
Phys.Rev. C {\bf 70}, 021901
  (2004) [hep-ph/0307267];
%
%\bibitem{BLR04}
G.E.~Brown, Ch,-H.~Lee, and M.~Rho,
%A new state of matter at high temperature as "sticky molasses",
hep-ph/0402207;
%
%\bibitem{Cassing05}
%THE HOT NON-PERTURBATIVE GLUON PLASMA IS AN ALMOST IDEAL COLORED LIQUID.
A. Peshier and W. Cassing, hep-ph/0502138.
%
\bibitem{Ellis}
G.W. Carter, O. Scavenius, I.N. Mishustin, and P.J. Ellis, Phys.
Rev. C  {\bf 61}, 045206 (2000).
%
\end{thebibliography}
\end{document}